\documentclass[aps,prd,preprintnumbers,groupedaddress,nofootinbib,amssymb,eqsecnum,notitlepage]{revtex4-1}
\usepackage{here}
\usepackage[dvipdfmx]{graphicx}
\usepackage{amsmath,amsthm,amssymb}
\usepackage{bm}
\usepackage{color}

\usepackage{amsfonts}
\usepackage{dcolumn}
\usepackage{hyperref}
\allowdisplaybreaks[1]
\usepackage{stackengine}

\begin{document}
\newcommand{\newc}{\newcommand}

\newc{\rk}[1]{{\color{red} #1}}
\newc{\ben}{\begin{eqnarray}}
\newc{\een}{\end{eqnarray}}
\newc{\be}{\begin{equation}}
\newc{\ee}{\end{equation}}
\newc{\ba}{\begin{eqnarray}}
\newc{\ea}{\end{eqnarray}}
\newc{\D}{\partial}
\newc{\rH}{{\rm H}}
\newc{\vp}{\varphi}
\newc{\rd}{{\rm d}}
\newc{\pa}{\partial}
\newc{\Mpl}{M_{\rm Pl}}

\newcommand{\ma}[1]{\textcolor{magenta}{#1}}
\newcommand{\cy}[1]{\textcolor{cyan}{#1}}
\newcommand{\mm}[1]{\textcolor{blue}{[MM:~#1]}}
\newcommand{\re}[1]{\textcolor{red}{#1}}

\begin{flushright}
WUCG-22-09 \\
\end{flushright}

\title{Inspiral gravitational waveforms from compact binary systems in Horndeski gravity}

\author{Yurika Higashino and Shinji Tsujikawa}

\affiliation{
Department of Physics, Waseda University, Shinjuku, Tokyo 169-8555, Japan}

\begin{abstract}

In a subclass of Horndeski theories with the speed of gravity equivalent to that of light, 
we study gravitational radiation emitted during the inspiral phase of compact binary systems. 
We compute the waveform of scalar perturbations under a post-Newtonian expansion 
of energy-momentum tensors of point-like particles that depend on 
a scalar field. This scalar mode not only gives rise to breathing and longitudinal 
polarizations of gravitational waves, but it is also responsible for 
scalar gravitational radiation in addition to energy loss associated with
transverse and traceless tensor polarizations.
We calculate the Fourier-transformed gravitational waveform of two 
tensor polarizations under a stationary phase approximation and show 
that the resulting waveform reduces to the one in a parametrized 
post-Einsteinian (ppE) formalism. 
The ppE parameters are directly related to a scalar charge in the 
Einstein frame, whose existence is crucial to allow the deviation from 
General Relativity (GR). We apply our general framework to several concrete 
theories and show that a new theory of spontaneous scalarization with 
a higher-order scalar kinetic term leaves interesting deviations from GR 
that can be probed by the observations of gravitational waves emitted from 
neutron star-black hole binaries. If the scalar mass 
exceeds the order of typical orbital frequencies $\omega \simeq 10^{-13}$~eV, which is the case for a recently proposed scalarized neutron star with 
a self-interacting potential, the gravitational waveform practically 
reduces to that in GR. 

\end{abstract}

\date{\today}


\maketitle

\section{Introduction}
\label{introsec}

The discovery of gravitational waves (GWs) \cite{LIGOScientific:2016aoc}  
opened up a new window for probing the physics in strong gravity 
regimes \cite{Berti:2015itd,Barack:2018yly,Berti:2018cxi}. 
Up to now, there have been many observed events of the 
coalescence of two black holes (BHs) \cite{LIGOScientific:2018mvr,LIGOScientific:2020tif}. 
The merger of two neutron stars (NSs) was first detected 
as the event GW170817 \cite{LIGOScientific:2017vwq}, 
together with an electromagnetic counterpart \cite{Goldstein:2017mmi}. 
This latter event constrained the speed of gravity to be very close to 
that of light \cite{LIGOScientific:2017zic}. 
There was also a possible NS-BH coalescence event 
GW190426-152155, albeit its highest false 
alarm rate \cite{LIGOScientific:2020aai,Broekgaarden:2021iew,Li:2020pzv,Niu:2021nic}. 
The increasing sensitivity in future GW observations will allow one
to detect more promising events of the NS-BH coalescence.

General Relativity (GR) is a fundamental theory of gravity consistent 
with submillimeter laboratory tests \cite{Hoyle:2000cv,Adelberger:2003zx}, 
solar system constraints \cite{Will:2014kxa}, 
and binary pulsar measurements \cite{Taylor:1982zz,Stairs:2003eg,Antoniadis:2013pzd}. 
However, there are several cosmological problems like the origins 
of inflation, dark energy, and dark matter, which are difficult to be 
resolved within the framework of GR and standard model of particle 
physics \cite{Lyth:1998xn,Copeland:2006wr,Bertone:2004pz}. 
To address these problems, one usually introduces additional degrees 
of freedom beyond those appearing in GR (two tensor polarizations). 
The simplest example is a scalar field $\phi$ minimally or nonminimally 
coupled to gravity. If there is a nonminimal coupling with the Ricci 
scalar $R$ of the form $F(\phi)R$, the gravitational interaction is generally 
modified from that in GR. Brans-Dicke (BD) theories \cite{Brans:1961sx} 
with a scalar potential $V(\phi)$ belong to such a class, which accommodates 
$f(R)$ gravity \cite{Starobinsky:1980te} as a special case. 
The applications of BD theories and $f(R)$ gravity to inflation and dark energy have been extensively performed 
in the literature (see the reviews \cite{Sotiriou:2008rp,DeFelice:2010aj}). 

In BD theories, the NS can have a scalar hair through the 
coupling between the scalar field and matter mediated by gravity. 
On the other hand, the no-hair property of BHs was proven in 
BD theories \cite{Hawking:1972qk,Bekenstein:1995un,Sotiriou:2011dz}. 
This means that the binary system containing at least one 
scalarized NS may leave some signatures for the deviation from 
GR in inspiral gravitational waveforms.
In massless BD theories, Eardley \cite{Eardley1975} estimated the change 
of an orbital 
period induced by dipole gravitational radiation in a compact binary 
system. In the same theories, Will \cite{Will:1994fb} computed
gravitational waveforms radiated during an inspiral phase of the 
compact binary up to the Newtonian quadrupole 
order (see also 
Refs.~\cite{Shibata:1994qd,Harada:1996wt,Brunetti:1998cc,Berti:2004bd,Scharre:2001hn,Chatziioannou:2012rf} for related works). 
Under a so-called post-Newtonian (PN) 
approximation \cite{Blanchet:2013haa} based on 
slow velocities of the binary system relative to the speed of 
light $c$, the gravitational radiation was also calculated up to 
2PN \cite{Lang:2013fna,Mirshekari:2013vb,Lang:2014osa,Sennett:2016klh} 
and 2.5PN \cite{Bernard:2022noq} orders, with the equations of 
motion up to 3PN order \cite{Bernard:2018hta}.

In massless BD theories given by the Lagrangian 
$L=(1/2)\phi R+\omega_{\rm BD}X/\phi$, where 
$X=-(1/2)\nabla^{\mu}\phi \nabla_{\mu}\phi $ is 
a field kinetic term with the covariant derivative operator 
$\nabla^{\mu}$, the solar-system tests of gravity 
put a tight bound on the BD parameter, 
$\omega_{\rm BD}>4.0 \times 10^4$ \cite{Will:2014kxa}.
With this constraint, the deviation from GR in strong 
gravity regimes is limited to be small.
In other words, the GW measurements need to reach high 
sensitivities to distinguish between BD theories and GR from 
the observed gravitational waveform. 
If the scalar field has a potential $V(\phi)$ with a heavy mass inside 
a nonrelativistic star, while having a light mass outside the star, 
it is possible to suppress the propagation of fifth forces in the solar system 
through a so-called chameleon mechanism \cite{Khoury:2003aq,Khoury:2003rn}.
In such cases the constraint on $\omega_{\rm BD}$ is 
loosened \cite{Faulkner:2006ub,Hu:2007nk,Capozziello:2007eu,Tsujikawa:2008uc,Berti:2012bp}, 
so there is more freedom to 
probe signatures of the modification of 
gravity in strong gravity environments. 
The gravitational radiation and tensor waves emitted from 
a compact binary have been computed in massive 
BD theories \cite{Alsing:2011er,Berti:2012bp,Sagunski:2017nzb,Liu:2020moh} 
and screened modified gravity 
in the Einstein frame \cite{Zhang:2017srh,Liu:2018sia,Niu:2019ywx}.

In the presence of a nonminimal coupling of the form $F(\phi)R$, 
there is yet the other scenario dubbed spontaneous scalarization 
of NSs \cite{Damour:1993hw,Damour:1996ke} in which the modification of 
gravity manifests itself only on the strong gravity background. 
Provided that $F(\phi)$ contains 
even power-law functions of $\phi$, the theory admits the 
existence of a nonvanishing field branch besides the GR branch 
($\phi=0$). Since the Ricci scalar $R$ coupled to the scalar field 
is large inside the NS, 
the GR branch can be unstable to trigger 
tachyonic growth of $\phi$ toward the other nontrivial branch. 
For the nonminimal coupling $F(\phi)=e^{-\beta \phi^2/(2\Mpl^2)}$ 
chosen by Damour and Esposito-Farese \cite{Damour:1993hw}, 
spontaneous scalarization can occur for the coupling constant 
$\beta \leq -4.35$ \cite{Harada:1998ge,Novak:1998rk,Sotani:2004rq,Silva:2014fca,Barausse:2012da}. 
On the other hand, the presence of a scalar charge for the scalarized 
solution leads to an energy loss through dipolar gravitational radiation.
This results in time variation of the orbital period of binary systems.
Indeed, binary-pulsar observations put the bound 
$\beta \geq -4.5$ \cite{Freire:2012mg,Shao:2017gwu,Anderson:2019eay}, 
so the coupling constant $\beta$ is confined 
in a limited range. Since the gravitational waveform emitted from 
compact binaries containing a NS should give an independent  
constraint on the scalar charge, it remains to be seen how the future 
GW observations place the bound on $\beta$. 
In Ref.~\cite{Niu:2021nic}, the authors started to derive constraints 
on $\beta$ by using the possible NS-BH coalescence event GW190426-152155.

Theory of spontaneous scalarization does not belong to a framework of 
BD theories, but it can be accommodated as a generalized class of BD 
theories by promoting the BD parameter $\omega_{\rm BD}$ to a scalar-field dependent function $\omega(\phi)$. 
Indeed, the gravitational radiation and waveforms 
in theories given by the Lagrangian 
$L=(1/2)\phi R+\omega (\phi) X/\phi$
have been investigated in 
Refs.~\cite{Alsing:2011er,Yunes:2011aa,Berti:2018cxi,Liu:2020moh}. 
Such theories belong to a scheme of Horndeski theories \cite{Horndeski:1974wa}-- most general scalar-tensor theories 
with second-order Euler-Lagrange equations 
of motion \cite{Deffayet:2011gz,Kobayashi:2011nu,Charmousis:2011bf}.
The subclass of Horndeski theories with the speed of gravity 
equivalent to that of light is given by the Lagrangian 
$L=G_2(\phi, X)-G_3(\phi,X)\square \phi
+G_4(\phi)R$ \cite{Kobayashi:2011nu,DeFelice:2011bh,Kase:2018aps}, 
where $G_2$ and $G_3$ are functions of $\phi$ and 
$X$, and 
$G_4$ is a function of $\phi$. This class of theories 
automatically evades the observational bound on 
the speed of gravity constrained by the GW170817 
event \cite{LIGOScientific:2017zic}.

Theories with the Lagrangian $L=G_2(\phi,X)+G_4(\phi)R$ 
accommodate not only the generalized massive BD theories with 
$L=(1/2)\phi R+\omega (\phi) X/\phi-V(\phi)$ but also nonminimally coupled 
k-essence \cite{Armendariz-Picon:1999hyi,Chiba:1999ka,Armendariz-Picon:2000nqq,Arkani-Hamed:2003pdi,Piazza:2004df,terHaar:2020xxb,Bezares:2021dma} 
containing higher-order kinetic terms like 
$\mu_2 X^2$ in $G_2$. When spontaneous scalarization occurs inside 
the NS, higher-order derivatives can be as large as 
the linear kinetic term around the surface of star. 
Indeed, we will propose a new scenario of spontaneous scalarization 
where the scalar charge is reduced by an additional 
term $\mu_2 X^2$. This allows a possibility for alleviating the 
tension of the coupling constant $\beta$ mentioned above. 
The modified scalar-field solution also affects the gravitational waveform 
radiated from the binary system containing a NS. 
Thus, it is convenient to provide a general scheme for 
confronting such theories with future GW observations 
of the NS-BH or NS-NS coalescence. 

In this paper, we compute the gravitational waveform emitted during the 
inspiral phase of compact binary systems in a subclass of Horndeski 
theories given by the Lagrangian 
$L=G_2(\phi, X)-G_3(\phi,X)\square \phi+G_4(\phi)R$. 
For this purpose, we perform the PN expansion of 
a source energy-momentum tensor and neglect nonlinear derivative 
terms arising from the cubic coupling $G_3(X)\square \phi$ 
outside the source. This amounts to neglecting nonlinear 
Galileon-type self-interactions \cite{Nicolis:2008in,Deffayet:2009wt} 
relative to the linear kinetic term. 
Hence it is difficult to accommodate the case in which the field derivative is 
suppressed in the exterior region of NSs by the Vainshtein 
mechanism \cite{Vainshtein:1972sx,Burrage:2010rs,DeFelice:2011th,Kimura:2011dc,Koyama:2013paa}, unless some specific scaling 
methods \cite{McManus:2016kxu,Renevey:2020tvr,Renevey:2021tcz} 
are employed. However, if the Vainshtein radius $r_V$ is of the same 
order as the NS radius $r_s$ ($\sim 10$~km),
the PN analysis used for the derivation of solutions to 
scalar perturbations from $r=r_s$ to an observer does not lose its validity. 
In such a case, the field derivative and scalar charge can be 
suppressed by the Vainshtein screening inside the NS, 
analogous to the findings in Ref.~\cite{Chagoya:2014fza,Ogawa:2019gjc}. 
Thus, the gravitational waveform derived in this paper can be applied 
to the case $r_V \lesssim r_s$. 

This paper is organized as follows.
In Sec.~\ref{eomsec}, we review the field equations of motion 
and the matter action of two point-like sources
in the subclass of Horndeski theories. 
In Sec.~\ref{weaksec}, we perform the weak-field expansions of 
metric and scalar field to study the propagation of GWs from 
the binary system of a quasicircular orbit to the observer 
and derive solutions to tensor GWs.
In Sec.~\ref{scapersec}, we obtain the time-domain gravitational waveforms 
corresponding to two transverse and longitudinal tensor polarizations 
as well as breathing and longitudinal modes. 
In Sec.~\ref{waveformsec}, we study the energy loss induced by the 
GW emission and derive the Fourier-transformed gravitational 
waveforms by using a stationary phase approximation. 
We show that the resulting waveform reduces to the one 
in a parameterized post-Einsteinian (ppE) 
framework \cite{Yunes:2009ke,Cornish:2011ys,Chatziioannou:2012rf,Tahura:2018zuq} 
and derive the ppE parameters in our theory. 
In Sec.~\ref{contheory}, we apply our general results to several concrete 
theories and clarify the relations between ppE parameters and 
the scalar charge in the Einstein frame. 
In particular, we show that a new theory of spontaneous scalarization 
with the higher-order derivative term $\mu_2 X^2$ in $G_2$ allows an 
interesting possibility for reducing the scalar charge in comparison 
to the case $\mu_2=0$, whose property 
can be probed in future GW observations. 
Sec.~\ref{conclusion} is devoted conclusions.  

Throughout the paper, we use the metric signature $(-,+,+,+)$ 
and natural units $c=\hbar=1$, where $\hbar$ is the reduced 
Planck constant.

\section{Subclass of Horndeski theories and field equations 
of motion}
\label{eomsec}

We consider a subclass of Horndeski theories \cite{Horndeski:1974wa} 
given by the action 
\be
{\cal S}=
\int {\rm d}^4 x \sqrt{-g} 
\left[ G_2(\phi,X)-G_3(\phi,X) \square \phi
+G_4(\phi)R \right]+{\cal S}_m(g_{\mu \nu}, \Psi_m)\,,
\label{action}
\ee
where $g$ is the determinant of metric tensor $g_{\mu\nu}$, 
$X=-(1/2)\nabla^{\mu}\phi \nabla_{\mu}\phi$ is the kinetic term 
of a scalar field $\phi$, 
and $\square \equiv g^{\mu \nu} \nabla_{\mu} \nabla_{\nu}$ 
is the d'Alembertian. 
The action ${\cal S}_m$ contains matter fields $\Psi_m$ 
minimally coupled to gravity. 
In theories (\ref{action}), there are no 
asymptotically flat spherically symmetric and static BH 
solutions with a scalar hair \cite{Hawking:1972qk,Bekenstein:1995un,Sotiriou:2011dz,Graham:2014mda,Faraoni:2017ock,Minamitsuji:2022vbi}. 
On the other hand, NSs can have scalar charges in the presence 
of nonminimal couplings $G_4(\phi)R$ \cite{Eardley1975}. 

Our aim is to compute a gravitational waveform emitted from 
inspiraling compact binaries containing at least one NS.
If the NS has a scalar hair, the waveform is subject to 
modifications in comparison to GR.
This allows us to probe signatures for the possible 
existence of a scalar field nonminimally coupled to gravity.
In theories given by the action (\ref{action}), the speed of gravity 
on the cosmological background is identical to that of light \cite{Kobayashi:2011nu,DeFelice:2011bh,Kase:2018aps}. 
We note that the equivalence principle can be generally 
violated in scalar-tensor theories including the action (\ref{action}). 
However, in concrete models discussed in Sec.~\ref{contheory}, 
we are interested in the case where the fifth force induced by 
scalar-gravitational couplings is suppressed on weak gravity 
backgrounds for the consistency with solar-system constraints. 
For the derivation of gravitational waveforms, we do not restrict 
the analysis to some specific models 
by the end of Sec.~\ref{waveformsec}.

We deal with the binary system of a quasicircular orbit as a collection 
of two point-like particles with masses $m_I(\phi)$, 
where $I=A, B$ for each particle. Since the existence of $\phi$ 
affects matter through gravitational field equations of motion, there is the 
$\phi$-dependence in $m_I$. 
The matter sector is expressed by the action \cite{Eardley1975}
\be
{\cal S}_m=-\sum_{I=A,B} \int m_I(\phi) {\rm d} \tau_I\,,
\label{matter}
\ee
where $\tau_I$ is the proper time 
along the world line $x_{I}^{\mu}$ of particle $I$.
The infinitesimal line element is given by 
\be
{\rm d}s^2=-{\rm d} \tau^2=g_{\mu \nu} 
{\rm d}x^{\mu} {\rm d}x^{\nu}\,.
\ee
Then, the matter action (\ref{matter}) can be written as
\be
{\cal S}_m=-\sum_{I=A,B} \int {\rm d}^4 x\,m_I(\phi) 
\sqrt{-g_{\mu \nu} 
{\rm d}x_I^{\mu} {\rm d}x_I^{\nu}}\,
\delta^{(4)} (x^{\mu}-x_I^{\mu})\,,
\label{matter2}
\ee
where $\delta^{(4)} (x^{\mu}-x_I^{\mu})$ is the four dimensional 
delta function. 
Varying (\ref{matter2}) with respect to 
$g_{\mu \nu}$, it follows that 
\be
\frac{\delta {\cal S}_m}{\delta g_{\mu \nu}}
=\sum_{I=A,B}\frac{1}{2} \int {\rm d}^4 x\, \int {\rm d}\tau_I\,m_I(\phi) 
u_I^{\mu} u_I^{\nu}\,\delta^{(4)} (x^{\mu}-x_I^{\mu})\,,
\label{Smva}
\ee
where $u_I^{\mu}={\rm d}x_I^{\mu}/{\rm d}\tau_I$ 
is the four velocity of particle $I$.
The matter energy-momentum tensor $T^{\mu \nu}$ 
is defined by 
\be
\delta {\cal S}_m=\frac{1}{2} \int {\rm d}^4 x \sqrt{-g}\, 
T^{\mu \nu}\delta g_{\mu \nu}\,.
\label{Smva3}
\ee
Comparing Eq.~(\ref{Smva3}) with Eq.~(\ref{Smva}), we obtain
\ba
T^{\mu \nu} &=&
\frac{1}{\sqrt{-g}} \sum_{I=A,B} \int {\rm d}\tau_I\,m_I(\phi) 
u_I^{\mu} u_I^{\nu}
\delta^{(4)} (x^{\mu}-x_I^{\mu}) \label{Tmunu0}\\
&=&
\frac{1}{\sqrt{-g}} 
\sum_{I=A,B} m_I(\phi) 
\frac{u_I^{\mu} u_I^{\nu}}{u_I^0}\delta^{(3)} ({\bm x}-{\bm x}_I(t))\,.
\label{Tmunu}
\ea
In the second line, we used ${\rm d}\tau_I={\rm d}x_I^{0}/u_I^{0}$ 
and integrated Eq.~(\ref{Tmunu0}) with respect to $x_I^{0}$. 
Note that $\delta^{(3)} ({\bm x}-{\bm x}_I(t))$ is a three dimensional 
delta function and that the time $t$ is determined by  
$t=x_I^0$. 

On using the property $g_{\mu \nu}u_I^{\mu} u_I^{\nu}=-1$, the trace 
of Eq.~(\ref{Tmunu}) yields
\be
T \equiv g_{\mu \nu}T^{\mu \nu}
=-\frac{1}{\sqrt{-g}} 
\sum_{I=A,B} m_I(\phi) 
\frac{1}{u_I^0}\delta^{(3)} ({\bm x}-{\bm x}_I(t))\,.
\label{trace}
\ee
On the other hand, the action (\ref{matter}) can be 
written in the form
\be
{\cal S}_m=-\sum_{I=A,B} \int {\rm d}^4 x\,m_I(\phi)
\frac{1}{u_I^0}\delta^{(3)} ({\bm x}-{\bm x}_I(t))\,.
\label{matterf}
\ee
Comparing Eq.~(\ref{trace}) with Eq.~(\ref{matterf}), 
it follows that 
\be
{\cal S}_m=\int {\rm d}^4 x \sqrt{-g}\,T(\phi)\,,
\label{Sm2}
\ee
whose form is used for the variation of ${\cal S}_m$ 
with respect to $\phi$.

Varying the action (\ref{action}) with respect to $g^{\mu\nu}$, 
we obtain the covariant gravitational field 
equations of motion \cite{Kobayashi:2011nu}
\ba
& &
-G_{2,X} \nabla_{\mu}\phi\nabla_{\nu}\phi-G_2 g_{\mu \nu}
+G_{3,X} \square \phi \nabla_{\mu}\phi\nabla_{\nu}\phi
+\nabla_{\mu}G_3 \nabla_{\nu}\phi
+\nabla_{\nu}G_3 \nabla_{\mu}\phi
-g_{\mu \nu} \nabla^{\lambda} G_3 \nabla_{\lambda} \phi \nonumber \\
& &
+2G_4 G_{\mu \nu}+2g_{\mu \nu} \left( 
G_{4,\phi} \square \phi-2X G_{4,\phi \phi} 
\right)-2G_{4,\phi} \nabla_{\mu} \nabla_{\nu}\phi
-2G_{4,\phi \phi}\nabla_{\mu}\phi\nabla_{\nu}\phi=T_{\mu \nu}\,,
\label{Ein0}
\ea
where we used the notations like $G_{2,X}=\partial G_2/\partial X$,  
$G_{4,\phi \phi}=\partial^2 G_4/\partial \phi^2$, etc.
The variation of (\ref{action}) with respect to $\phi$ leads to the scalar-field equation of motion 
\be
\nabla^{\mu} J_{\mu}={\cal P}_{\phi}\,,
\label{fieldeq}
\ee
where 
\ba
J_{\mu} &=& -G_{2,X} \nabla_{\mu} \phi+G_{3,X} \square \phi 
\nabla_{\mu}\phi+G_{3,X} \nabla_{\mu}X+2G_{3,\phi} \nabla_{\mu}\phi\,,\\
{\cal P}_{\phi} &=& G_{2,\phi}+\nabla^{\mu} G_{3,\phi} \nabla_{\mu}\phi
+G_{4,\phi}R+T_{,\phi}\,.
\label{Pphi}
\ea
The $\phi$ dependence in $T(\phi)$ influences the scalar-field equation 
through the last term in Eq.~(\ref{Pphi}). 
The Ricci scalar $R$ in ${\cal P}_{\phi}$ is affected by 
the presence of matter through the gravitational Eq.~(\ref{Ein0}).
Taking the trace of Eq.~(\ref{Ein0}), we obtain
\be
2G_4 R=2X G_{2,X}-4G_2-2X G_{3,X} \square \phi-2\nabla^{\mu} G_3 
\nabla_{\mu} \phi+6 \left( G_{4,\phi} \square \phi
-2XG_{4,\phi \phi} \right)-T\,.
\ee
Then, we can express ${\cal P}_{\phi}$ in the following form
\ba
{\cal P}_{\phi}
&=& G_{2,\phi}+\nabla^{\mu} G_{3,\phi} \nabla_{\mu}\phi
+\frac{G_{4,\phi}}{G_4} \left( X G_{2,X}-2G_2-XG_{3,X}
\square \phi-\nabla^{\mu} G_3 
\nabla_{\mu} \phi+3G_{4,\phi}\square \phi-6XG_{4,\phi \phi} \right)
\nonumber \\
& &
+T_{,\phi}-\frac{G_{4,\phi}}{2G_4}T\,.
\label{Pphi2}
\ea
The last two contributions to Eq.~(\ref{Pphi2}) work as matter 
source terms for the scalar-field equation.

The matter action (\ref{matter}) can be also expressed in the form  
\be
{\cal S}_m=-\sum_{I=A,B} \int m_I (\phi (x_I^\lambda)) 
\sqrt{-g_{\mu \nu} (x_I^\lambda) u_I^{\mu}u_I^{\nu}}\,{\rm d}\tau_I\,.
\ee
Varying this action with respect to $x_I^\lambda$ and integrating 
it by parts, we obtain 
\be
\delta {\cal S}_m=-\sum_{I=A,B} \int \left[ \frac{{\rm d}}{{\rm d}\tau} 
\left( m_I g_{\mu \lambda} u_I^{\mu} \right) 
-\frac{1}{2} m_I \frac{\partial g_{\mu \nu}}{\partial x_I^{\lambda}}
u_I^{\mu} u_I^{\nu}+m_{I,\phi} \frac{\partial \phi}{\partial x_I^{\lambda}}
\right] \delta x_I^{\lambda}{\rm d}\tau\,.
\label{SMf0}
\ee
Then, the equation of motion for the $I$-th particle is given by 
\be
\frac{{\rm d}}{{\rm d}\tau} 
\left( m_I g_{\mu \lambda} u_I^{\mu} \right) 
-\frac{1}{2} m_I \frac{\partial g_{\mu \nu}}{\partial x_I^{\lambda}}
u_I^{\mu} u_I^{\nu}+m_{I,\phi} \frac{\partial \phi}{\partial x_I^{\lambda}}
=0\,.
\label{SMf}
\ee
Multiplying Eq.~(\ref{SMf}) by $g^{\alpha \lambda}$ and using 
${\rm d}m_I/{\rm d}\tau=m_{I,\phi}u^{\beta} \nabla_{\beta}\phi$
and
${\rm d} g_{\mu \lambda}/{\rm d} \tau
=(\partial g_{\mu \lambda}/\partial x_I^{\nu})u_I^{\nu}/2
+(\partial g_{\lambda \mu}/\partial x_I^{\nu})u_I^{\nu}/2$, it follows that 
\be
m_I \left( \frac{{\rm d} u_I^{\alpha}}{{\rm d} \tau}+
\Gamma^{\alpha}_{\mu \nu}u_I^{\mu} u_I^{\nu} \right)
+m_{I,\phi} \left( \nabla^{\alpha}\phi+u_I^{\alpha} u_I^{\beta} 
\nabla_{\beta}\phi \right)=0\,,
\label{geo}
\ee
where $\Gamma^{\alpha}_{\mu \nu}$ is the Christoffel  symbol.
The $\phi$ dependence in $m_I$ modifies the standard geodesic equation. 
One can express Eq.~(\ref{geo}) in a simple form \cite{Eardley1975}
\be
u_I^{\beta} \nabla_{\beta} \left( m_I u_I^{\alpha} \right)
=-m_{I,\phi}\nabla^{\alpha}\phi\,.
\label{uA}
\ee
In terms of the matter energy-momentum tensor given by Eq.~(\ref{Tmunu0}), 
Eq.~(\ref{uA}) is equivalent to the continuity equation
$\nabla_{\beta} T^{\alpha \beta}=T_{,\phi} \nabla^{\alpha} \phi$. 
This latter equation coincides with the one derived in Ref.~\cite{Will:1989sk}
in BD theories.

Equations (\ref{Ein0}), (\ref{fieldeq}), and (\ref{geo}) are the master equations 
used to describe the dynamics of gravity, scalar-field, and point-like 
particles, respectively.

\section{Weak field expansion}
\label{weaksec}

To compute the gravitational waveform emitted from the inspiraling compact 
binary, we need to study the propagation of GWs from the binary to an observer. For this purpose, 
we expand the metric $g_{\mu \nu}$ about a Minkowski background 
and the scalar field $\phi$ around a constant asymptotic cosmological 
value $\phi_0$, as \cite{Will:1994fb}
\be
g_{\mu \nu}=\eta_{\mu \nu}+h_{\mu \nu}\,,\qquad 
\phi=\phi_0+\varphi\,,
\ee
where $\eta_{\mu \nu}={\rm diag} (-1,1,1,1)$, and  
$h_{\mu \nu}$ and $\varphi$ are the perturbed quantities.
We would like to calculate the gravitational waveform 
associated with $h_{\mu \nu}$ and $\varphi$ 
up to quadrupole order. 
We perform the expansions of 
Eqs.~(\ref{Ein0}) and (\ref{fieldeq}) with respect to the perturbations 
$h_{\mu \nu}$ and $\varphi$ and derive a quadrupole formlula 
for tensor GWs in this section.

The scalar-field equation (\ref{Pphi2}) contains the matter source 
term $T_{,\phi}-G_{4,\phi}T/(2G_4)$. 
The trace $T$ acquires the $\phi$ dependence through the mass 
term $m_I(\phi)$ in Eq.~(\ref{trace}). 
We expand $m_I(\phi)$ and $G_4(\phi)$ around the background 
field value $\phi=\phi_0$, respectively, as
\ba
m_I(\phi) &=&
m_I(\phi_0) \left[ 1+\alpha_I 
\left( \frac{\varphi}{\Mpl} \right)+\frac{1}{2} \left( \alpha_I^2
+\beta_I \right) \left( \frac{\varphi}{\Mpl} \right)^2 
\right]\,,
\label{mAexpan}\\
G_4(\phi) &=&
G_4(\phi_0) \left[ 1+g_4 
\left( \frac{\varphi}{\Mpl} \right)+\frac{1}{2} \left( 
g_4^2 +\gamma_4 \right) \left( \frac{\varphi}{\Mpl} \right)^2 
\right]\,,
\label{G4expan}
\ea
where $\Mpl=2.4354 \times 10^{18}$~GeV is 
the reduced Planck mass, and 
\ba
\alpha_I &\equiv&
\Mpl \frac{{\rm d} \ln m_I(\phi)}
{{\rm d} \phi}\biggr|_{\phi=\phi_0}\,,\qquad 
\beta_I \equiv \Mpl^2 \frac{{\rm d}^2 \ln m_I(\phi)}
{{\rm d}\phi^2}\biggr|_{\phi=\phi_0}\,,\label{alphaIJ}\\
g_4 &\equiv& \Mpl \frac{{\rm d} \ln G_4(\phi)}
{{\rm d} \phi}\biggr|_{\phi=\phi_0}\,,\qquad 
\gamma_4 \equiv \Mpl^2 \frac{{\rm d}^2 \ln G_4(\phi)}
{{\rm d}\phi^2}\biggr|_{\phi=\phi_0}\,.
\label{g4J}
\ea
On using Eq.~(\ref{trace}), the matter source terms in 
Eq.~(\ref{Pphi2}) evaluated on the Minkowski background 
are expressed as
\be
T_{,\phi}-\frac{G_{4,\phi}}{2G_4}T\biggr|_{\phi=\phi_0}=
-\frac{1}{\Mpl} \sum_{I=A,B} \hat{\alpha}_I
m_I(\phi_0)
\frac{1}{u_I^0} \delta^{(3)} ({\bm x}-{\bm x}_I(t))\,,
\label{Tphi}
\ee
where 
\be
\hat{\alpha}_I \equiv \alpha_I -\frac{g_4}{2}\,.
\label{halI}
\ee
As we will show in Sec.~\ref{contheory}, the quantity  
$\hat{\alpha}_I$ is directly related to a scalar charge. 
In this sense, $\hat{\alpha}_I$ is a more fundamental 
physical quantity than $\alpha_I$. 
It is known that the theory (\ref{action}) does not have 
hairy BH solutions, in which case $\hat{\alpha}_I=0$. 
On the other hand, the NS can have scalar hairs, in which case 
$\hat{\alpha}_I \neq 0$. 
As we will see in Sec.~\ref{waveformsec}, the scalar charge 
$\hat{\alpha}_I$ appears in the gravitational waveform as 
a quantity charactering the deviation from GR. 
One can also consider the following 
sensitivity parameter \cite{Eardley1975}
\be
s_I \equiv \frac{{\rm d} \ln m_I(\phi)}
{{\rm d} \ln \phi}\biggr|_{\phi=\phi_0}
=\frac{\phi_0}{\Mpl} \alpha_I\,.
\label{sI}
\ee
In BD theories given by $G_4=\phi/(16 \pi)$, 
we have $g_4=\Mpl/\phi_0$ and hence 
$\hat{\alpha}_I=(\Mpl/\phi_0)(s_I-1/2)$. 
Then, the no-hair BH in BD theories corresponds to the 
sensitivity parameter $s_I=1/2$. 
Depending on the theories under consideration, however, 
$s_I$ defined in Eq.~(\ref{sI}) can be affected by an ambiguity of the asymptotic 
value of $\phi_0$.
In the following we will use $\hat{\alpha}_I$ 
instead of $s_I$, as the former has a direct relation 
with the scalar charge. 

\subsection{Perturbation equations up to second order}

We expand the field equations of motion up to second order in metric and 
scalar-field perturbations. 
In doing so, we use the properties 
$\nabla_{\mu} \phi=\partial_{\mu} \varphi$ and 
$\delta X=-\eta^{\mu \nu} \partial_{\mu} \varphi \partial_{\nu} \varphi/2$ up to 
quadratic order,  
where $ \partial_{\mu} \equiv \partial/ \partial x^{\mu}$. 
We also exploit the following relation 
\be
\square \phi=(\eta^{\mu \nu}-h^{\mu \nu}) \nabla_{\mu} \partial_{\nu} \phi
=\square_{\rm M} \vp-h^{\mu \nu} \partial_{\mu} \partial_{\nu}\vp
- \left( \pa_{\mu} h^{\mu \alpha}-\frac{1}{2} 
\eta^{\alpha \beta} \pa_{\beta} h \right)\partial_{\alpha} \vp
+{\cal O}(\varepsilon^3)\,,
\label{squarephi}
\ee
where ${\cal O}(\varepsilon^3)$ means the third-order 
perturbations, and 
\be
\square_{\rm M} \equiv \eta^{\mu \nu} \pa_{\mu} \pa_{\nu}
=-\frac{\pa^2}{\pa t^2}+\nabla^2\,,
\ee
with $\nabla^2$ being the three dimensional Laplacian 
in Minkowski spacetime. 
Expanding Eqs.~(\ref{Ein0}) and (\ref{fieldeq}) up to second order 
in perturbations, it follows that \cite{Hou:2017cjy}
\ba
& &
-G_2 \eta_{\mu \nu}-G_2 h_{\mu \nu}-G_{2,\phi}\vp \eta_{\mu \nu}
-G_{2,\phi} \vp h_{\mu \nu}+\frac12 G_{2,X} \pa^{\mu} \vp 
\pa_{\mu} \vp\,\eta_{\mu \nu}-\frac12 G_{2,\phi \phi} \vp^2 \eta_{\mu \nu}
-G_{2,X} \pa_{\mu} \vp \pa_{\nu} \vp \nonumber \\
& &+2 G_{3,\phi} \pa_{\mu} \vp \pa_{\nu} \vp
-\eta_{\mu \nu}G_{3,\phi} \pa^{\lambda} \vp \pa_{\lambda} \vp
+2(G_4+G_{4,\phi}\vp) \delta G_{\mu \nu}^{(1)}
+2G_4 \delta G_{\mu \nu}^{(2)}
+2h_{\mu \nu} G_{4,\phi} \square_{\rm M} \vp
-2\eta_{\mu \nu} G_{4,\phi} h^{\alpha \beta} 
\pa_{\alpha} \pa_{\beta}\vp 
\nonumber \\
& &
+2\eta_{\mu \nu} \left( G_{4,\phi} \square_{\rm M} \vp
+G_{4,\phi \phi} \vp \square_{\rm M} \vp+G_{4,\phi \phi} 
\partial^{\lambda} \vp \partial_{\lambda} \vp \right)
-\eta_{\mu \nu} G_{4,\phi} \pa_{\alpha}\vp
\left( 2\pa_{\beta}h^{\alpha \beta}-\eta^{\alpha \beta} 
\pa_{\beta} h \right)-2G_{4,\phi}\pa_{\mu} \pa_{\nu}\vp
\nonumber \\
& & 
+G_{4,\phi} \left( 2\pa_{\mu} {h_{\nu}}^{\alpha}
-\eta^{\alpha \beta} \pa_{\beta} h_{\mu \nu} \right) 
\pa_{\alpha}\vp -2G_{4,\phi \phi} \vp \pa_{\mu} \pa_{\nu} \vp
-2G_{4,\phi \phi} \pa_{\mu} \vp \pa_{\nu} \vp=
T_{\mu \nu}^{(1)}+T_{\mu \nu}^{(2)}\,,\label{per1}\\
& &
\left( G_{2,X}-2G_{3,\phi} \right)
\left[
\square_{\rm M} \vp-h^{\mu \nu} \partial_{\mu} \partial_{\nu}\vp
-\left( \pa_{\mu} h^{\mu \alpha}-\frac{1}{2} 
\eta^{\alpha \beta} \pa_{\beta} h \right)\partial_{\alpha} \vp \right]
+\left( G_{2,\phi X}-2G_{3,\phi \phi} \right) 
\left( \vp \square_{\rm M} \vp+\pa^{\mu} \vp \pa_{\mu} \vp \right) 
\nonumber \\
& & -G_{3,X} \left[ (\square_{\rm M} \vp)^2
-\pa^{\mu} \pa^{\nu} \vp\pa_{\mu} \pa_{\nu} \vp \right]
+G_{2,\phi}+G_{2,\phi \phi} \vp+\frac{1}{2} G_{2,\phi \phi \phi} \vp^2
-\left( \frac{1}{2} G_{2,\phi X}-2G_{3,\phi \phi} \right) 
\pa^{\mu} \vp \pa_{\mu} \vp\nonumber \\
& &
+\left( G_{4,\phi}+G_{4,\phi \phi} \vp \right) \delta R^{(1)}
+G_{4,\phi} \delta R^{(2)}=-T_{,\phi}^{(1)}-T_{,\phi}^{(2)}\,,\label{per2}
\ea
where $\delta G_{\mu \nu}$ and $\delta R$ are the perturbed Einstein tensor and 
Ricci scalar, respectively, $h\equiv \eta^{\mu \nu} h_{\mu \nu}$ is the trace of 
$h_{\mu \nu}$, and the upper subscripts ``$(1)$'' and ``$(2)$'' represent 
the first- and second-order perturbations, respectively.
The explicit forms of $\delta G_{\mu \nu}^{(1)}$ and $\delta R^{(1)}$ 
are given, respectively, by 
\ba
\delta G_{\mu \nu}^{(1)} &=&
-\frac{1}{2}
\left( \square_{\rm M}h_{\mu \nu}
-\eta_{\mu \nu} \square_{\rm M}h-2 \pa_{\mu} \pa^{\alpha} 
h_{\nu \alpha}+\pa_{\mu} \pa_{\nu}h+\eta_{\mu \nu} \pa_{\alpha} 
\pa_{\beta} h^{\alpha \beta} \right)\,,\\
\delta R^{(1)} &=&
\pa_{\mu} \pa_{\nu} h^{\mu \nu}-\square_{\rm M}h\,.
\ea
In Eqs.~(\ref{per1}) and (\ref{per2}), the quantities $G_{2,3,4}$ and their 
$\phi$, $X$ derivatives should be evaluated on the Minkowski background 
with the field value $\phi=\phi_0$, e.g., $G_4=G_4(\phi_0)$. 
For the consistency of Eq.~(\ref{per1}), the background term $-G_2 \eta_{\mu \nu}$ 
needs to vanish. Similarly, the term $G_{2,\phi}$ in Eq.~(\ref{per2}) must vanish, 
so that 
\be
G_2(\phi_0)=0\,,\qquad G_{2,\phi}(\phi_0)=0\,.
\ee

We introduce the following quantity
\be
\theta_{\mu \nu} \equiv h_{\mu \nu}-\frac{1}{2} \eta_{\mu \nu}h-\eta_{\mu \nu} 
g_4 \frac{\vp}{\Mpl}\,.
\label{theta}
\ee
Taking the trace of Eq.~(\ref{theta}) and defining $\theta \equiv 
\eta^{\mu \nu}\theta_{\mu \nu}$, we find 
\be
h=-\theta-4g_4 \frac{\vp}{\Mpl}\,,
\label{hth1}
\ee
so that $h_{\mu \nu}$ is expressed as 
\be
h_{\mu \nu}=\theta_{\mu \nu}-\frac{1}{2} \eta_{\mu \nu} 
\theta-\eta_{\mu \nu} 
g_4 \frac{\vp}{\Mpl}\,.
\label{hth2}
\ee
Substituting Eqs.~(\ref{hth1}) and (\ref{hth2}) into 
Eq.~(\ref{per1}), we obtain 
\be
-\square_{\rm M} \theta_{\mu \nu}+2\pa_{\mu} \pa^{\alpha} 
\theta_{\nu \alpha}-\eta_{\mu \nu} \pa^{\alpha} \pa^{\beta} 
\theta_{\alpha \beta}=\tau_{\mu \nu}\,,
\label{thetamn}
\ee
where 
\be
\tau_{\mu \nu}=\frac{T_{\mu \nu}^{(1)}}{G_4}
+{\cal O} \left(\theta^2, \vp^2, \theta \vp, T_{\mu \nu}^{(2)},\cdots 
\right)\,.
\label{taumunu}
\ee
In the following, we choose the Lorentz gauge condition 
\be
\pa^{\nu} \theta_{\mu \nu}=0\,.
\ee
Under this gauge choice, Eq.~(\ref{thetamn}) is simplified to 
\be
\square_{\rm M} \theta_{\mu \nu}=-\tau_{\mu \nu}\,,
\label{thetaf}
\ee
with $\tau_{\mu \nu}$ satisfying 
\be
\pa^{\nu} \tau_{\mu \nu}=0\,.
\label{contau}
\ee
We note that the leading-order contribution to $\tau_{\mu \nu}$ 
is the first-order perturbation $T_{\mu \nu}^{(1)}/G_4(\phi_0)$.

\subsection{Linear perturbations and quadrupole formula of 
tensor waves}
\label{linearso}

Let us first derive the solutions to $h_{\mu \nu}$ and $\varphi$ 
at linear order. 
At first order in perturbations, Eqs.~(\ref{per1}) and (\ref{per2})
reduce, respectively, to 
\ba
& &
2G_4 \delta G_{\mu \nu}^{(1)}+
2\eta_{\mu \nu} G_{4,\phi} \square_{\rm M} \vp
-2G_{4,\phi}\pa_{\mu} \pa_{\nu}\vp=T_{\mu \nu}^{(1)}\,,
\label{pereq1}\\
& &
\left( G_{2,X}-2G_{3,\phi} \right) \square_{\rm M} \vp
+G_{2,\phi \phi} \vp+G_{4,\phi} \delta R^{(1)}=-T_{,\phi}^{(1)}\,.
\label{pereq2}
\ea
Taking the trace of Eq.~(\ref{pereq1}), defining 
$T^{(1)} \equiv \eta^{\mu \nu} T_{\mu \nu}^{(1)}$, and 
using the property $\eta^{\mu \nu} \delta G_{\mu \nu}^{(1)}
=-\delta R^{(1)}$, we obtain
\be
\delta R^{(1)}=\frac{6G_{4,\phi}\square_{\rm M}\vp-T^{(1)}}
{2G_4}\,.
\label{delR}
\ee
Substituting Eq.~(\ref{delR}) into Eq.~(\ref{pereq2}), 
we find
\be
\left( \square_{\rm M}-m_s^2 \right) \vp
=-\frac{1}{\zeta_0} \left[ T_{,\phi}^{(1)}
-\frac{g_4}{2 \Mpl}T^{(1)} \right]\,,
\label{vpli}
\ee
where 
\be
\zeta_0 \equiv G_{2,X}-2G_{3,\phi}
+\frac{3G_{4,\phi}^2}{G_4} \biggr|_{\phi=\phi_0}\,,
\qquad
m_s^2 \equiv -\frac{G_{2,\phi \phi}(\phi_0)}{\zeta_0}\,.
\label{ze0def}
\ee
The quantity $m_s$ corresponds to the mass of the scalar field. 
In the presence of a scalar potential $V(\phi)$ appearing as 
the term $-V(\phi)$ in $G_2(\phi,X)$, 
the mass squared is given by $V_{,\phi \phi}/\zeta_0$. 
{}From Eqs.~(\ref{taumunu}) and (\ref{thetaf}), 
the linear-order perturbation $\theta_{\mu \nu}$ obeys 
\be
\square_{\rm M} \theta_{\mu \nu}=-\frac{T_{\mu \nu}^{(1)}}{G_4(\phi_0)}\,.
\label{graeq}
\ee
To solve Eq.~(\ref{graeq}) for $\theta_{\mu \nu}$, we consider the 
Green function satisfying $\square_{\rm M}\,{\cal G} (x-x')=\delta^{(4)} (x-x')$, 
where $x$ represents the four dimensional coordinate $x^{\mu}=(t, {\bm x})$ and $\delta^{(4)}(x)$ is the four dimensional delta function. 
We exploit the fact that the integrated solution to this equation is  
expressed in the form  
\be
{\cal G} (x-x')=-\frac{1}{4\pi |{\bm x}-{\bm x}'|}
\delta (t_{\rm ret}-t')\,,
\ee
where $t_{\rm ret}=t-|{\bm x}-{\bm x}'|$ is the retarded time.
Then, the solution to Eq.~(\ref{graeq}) at spacetime point $x$ is 
given by 
\be
\theta_{\mu \nu} (x)=
\frac{1}{4\pi G_4(\phi_0)} \int \rd^4 x' \frac{\delta (t_{\rm ret}-t')}
{|{\bm x}-{\bm x}'|} T_{\mu \nu}^{(1)}(x')
=\frac{1}{4\pi G_4(\phi_0)} \int \rd^3 x' 
\frac{T_{\mu \nu}^{(1)}(t-|{\bm x}-{\bm x}'|, {\bm x}')}
{|{\bm x}-{\bm x}'|}\,.
\label{thetax}
\ee
Since the metric components $\theta_{0 \mu}$ do not correspond to 
the dynamical degrees of freedom of GWs, we will study the propagation 
of spatial components $\theta_{ij}$ in the following discussion. 
We would like to compute $\theta_{ij}(x)$ at an observer 
position ${\bm x}=D{\bm n}$ far away from a binary source, 
where ${\bm n}$ is a unit vector. 
For $|{\bm x}'|$ at most of order a typical radius of the source $d$, 
we have $|{\bm x}-{\bm x}'|=D-{\bm x}' \cdot {\bm n}+{\cal O}(d^2/D)$ 
and hence $t-|{\bm x}-{\bm x}'| \simeq t-D+{\bm x}' \cdot {\bm n}$.
Provided that typical velocities of the source are much smaller than 
the speed of light, we can expand $T_{\mu \nu}(t-|{\bm x}-{\bm x}'|, {\bm x}')$ 
about the retarded time $t-D$. 
For $D \gg d$, the denominator of Eq.~(\ref{thetax}) can be 
approximated as $|{\bm x}-{\bm x}'| \simeq D$. Then, it follows that 
\be
\theta_{ij} (x)=\frac{1}{4\pi G_4(\phi_0) D} \sum_{\ell=0}^{\infty} 
\frac{1}{\ell !} \frac{\pa^\ell}{\pa t^\ell} \int \rd^3 x'\, 
T_{ij}^{(1)} \left( t-D, {\bm x}' \right) 
\left( {\bm x}' \cdot {\bm n} \right)^\ell\,.
\label{thetaij}
\ee
On using the continuity equation $\pa^{\nu}T_{\mu \nu}^{(1)}=0$ 
arising from Eq.~(\ref{contau}), there is the 
relation \cite{Maggiore:2007ulw}
\be
\int \rd^3 x'\, T_{ij}^{(1)}(t, {\bm x}')
=\frac{1}{2} \frac{\pa^2}{\pa t^2}\int \rd^3 x'\, T_{00}^{(1)}(t, {\bm x}')
x_i' x_j'\,.
\label{Tijre}
\ee
Then, the leading-order term of Eq.~(\ref{thetaij}) (i.e., $\ell=0$) yields
\be
\theta_{ij} (x)=\frac{1}{8\pi G_4(\phi_0) D} 
\frac{\pa^2}{\pa t^2} \int \rd^3 x'\, 
T_{00}^{(1)} \left( t-D, {\bm x}' \right) x_i' x_j'\,.
\label{thetaij2}
\ee
Under the low-velocity approximation of point sources, 
the leading-order contribution to the (00) component 
of Eq.~(\ref{Tmunu}) on the Minkowski background is given by 
\be
{T^{00}}^{(1)}(x)=\sum_{I=A,B} m_I
\delta^{(3)} ({\bm x}-{\bm x}_I(t))\,.
\label{T00up}
\ee
{}From Eqs.~(\ref{thetaij2}) and (\ref{T00up}), we obtain 
the following quadrupole formula
\be
\theta^{ij} (x)=\frac{1}{8\pi G_4(\phi_0) D} 
\frac{\pa^2}{\pa t^2} \sum_{I=A,B} m_I
x_I^i x_I^j\,.
\label{thetaij3}
\ee
Since $\theta^{ij} (x)$ depends on the motion of sources, 
we derive the geodesic equations of motion at Newtonian order 
in Sec.~\ref{geosec}.

\subsection{Geodesic equations at Newtonian order}
\label{geosec}

Under the approximation that the typical velocities of sources are much 
smaller than the speed of light ($|u_I^i| \ll 1$ with $\tau \simeq t$), 
the spatial components of Eq.~(\ref{geo}) translate to
\be
\frac{\rd^2 x_I^i}{\rd t^2}-\frac{1}{2} \pa^i h_{00}
+\alpha_I \frac{\pa^i \vp}{\Mpl}=0\,,
\label{geoeq1}
\ee
where we used $\Gamma^{i}_{00} \simeq -\pa^i h_{00}/2$. 
The particle motion is affected by the spatial derivatives of 
$h_{00}$ and $\vp$, so we compute these terms in the following.
At leading order in the PN approximation, 
the only nonvanishing component of 
$T_{\mu \nu}$ is given by 
\be
T_{00}=T_{00}^{(1)}=\sum_{I=A,B}m_I \delta^{(3)} \left(
{\bm x}-{\bm x}_I \right)\,.
\ee
In the Newtonian limit the background spacetime is stationary, so 
the $(00)$ component of Eq.~(\ref{graeq}) yields
\be
\nabla^2 \theta_{00}({\bm x})=-\frac{1}{G_4(\phi_0)}\sum_{I=A,B}
m_I\delta^{(3)} \left(
{\bm x}-{\bm x}_I \right)\,.
\ee
This is integrated to give
\be
\theta_{00}({\bm x})=\frac{U({\bm x})}{4\pi G_4(\phi_0)}\,,
\label{theta00}
\ee
where 
\be
U({\bm x}) \equiv \sum_{I=A,B}\frac{m_I}{|{\bm x}-{\bm x}_I|}\,, 
\ee
with the trace $\theta({\bm x})=-U({\bm x})/[4\pi G_4(\phi_0)]$.

At linear order, the scalar-field perturbation obeys 
Eq.~(\ref{vpli}) with $T^{(1)}=-T_{00}^{(1)}$. 
In the stationary Newtonian limit, this equation yields
\be
\left( \nabla^2-m_s^2 \right) \vp({\bm x})
=\sum_{I=A,B} \frac{\hat{\alpha}_I m_I}{\zeta_0 \Mpl} 
\delta^{(3)} \left( {\bm x}-{\bm x}_A \right)\,,
\label{vpli1}
\ee
which is integrated to give 
\be
\vp({\bm x})=\frac{U_s({\bm x})}{8\pi \zeta_0 \Mpl}\,,
\label{vp2}
\ee
where 
\be
U_s({\bm x}) \equiv
-2 \sum_{I=A,B} \hat{\alpha}_I 
m_I \frac{e^{-m_s|{\bm x}-{\bm x}_I|}}{|{\bm x}-{\bm x}_I|}\,.
\ee
Substituting Eqs.~(\ref{theta00}) and (\ref{vp2}) into 
Eq.~(\ref{hth2}), the leading-order components of $h_{\mu \nu}$ 
are given by 
\be
h_{00}=\frac{1}{8\pi} \left[ \frac{U}{G_4(\phi_0)}
+\frac{g_4 U_s}{\zeta_0 \Mpl^2} \right]\,,\qquad 
h_{0i}=0\,,\qquad 
h_{ij}=\frac{1}{8\pi} \left[ \frac{U}{G_4(\phi_0)}
-\frac{g_4 U_s}{\zeta_0 \Mpl^2} \right] \delta_{ij}\,.
\label{h00}
\ee
On using the solutions (\ref{vp2}) and (\ref{h00}) for a binary system,
the equations of motion of particles $A$ and $B$ following from 
Eq.~(\ref{geoeq1}) are
\be
\frac{\rd^2 x_A^i}{\rd t^2}=
-\frac{ \tilde{G} m_B r^i}{r^3} \,,\qquad 
\frac{\rd^2 x_B^i}{\rd t^2}=
\frac{ \tilde{G} m_A r^i}{r^3} \,,
\label{geA}
\ee
where $r^i=x_A^i-x_B^i$, $r=|r^i|$, and 
\be
\tilde{G} \equiv \frac{1}{16 \pi G_4(\phi_0)} 
\left[
1+\frac{4G_4(\phi_0)}{\zeta_0 \Mpl^2} 
\hat{\alpha}_A \hat{\alpha}_B
\left( 1+m_s r \right) e^{-m_s r} 
\right]\,.
\label{q}
\ee
The quantity $\tilde{G}$ corresponds to an effective gravitational 
coupling between two point-like particles, which contains 
the product of $\hat{\alpha}_A$ and $\hat{\alpha}_B$. 
The relative displacement of two sources obeys the differential equation 
\be
\mu \ddot{r}^i=
-\frac{\tilde{G} m_A m_B}{r^3} r^i\,,
\label{gere}
\ee
where a dot represents the derivative with respect to $t$, 
and $\mu$ is the reduced mass defined by 
\be
\mu \equiv \frac{m_A m_B}{m_A+m_B}\,.
\label{mudefi}
\ee
\subsection{Tensor waves at quadrupole order 
from a quasicircular orbit}
\label{grarasec}

For a quasicircular orbit of a binary system, we will simplify the 
quadrupole formula (\ref{thetaij3}). 
Introducing the center of mass 
\be
x^i_{\rm CM}=\frac{m_A x^i_A+m_B x^i_B}{m}\,,\qquad
{\rm with} \qquad m \equiv m_A+m_B\,,
\label{massAB}
\ee
we can express Eq.~(\ref{thetaij3}) in the form 
\be
\theta^{ij} (x)=\frac{1}{8\pi G_4(\phi_0) D} 
\frac{\pa^2}{\pa t^2} 
\left( m x_{\rm CM}^i x_{\rm CM}^j
+\mu r^i r^j \right)\,,
\label{thetaij4}
\ee
For an isolated binary system the center of mass is not accelerating, so 
it does not contribute to the generation of GWs.
Then, we can choose the frame $x_{\rm CM}^i=0$, i.e., 
\be
m_A x^i_A+m_B x^i_B=0\,,
\ee
without loss of generality. 
Then, Eq.~(\ref{thetaij4}) reduces to 
\be
\theta^{ij} (x)=\frac{\mu}{8\pi G_4(\phi_0) D} 
\left( 2 \dot{r}^i \dot{r}^j+
\ddot{r}^i r^j+ r^i \ddot{r}^j \right)\,.
\label{thetaijx}
\ee
Substituting Eq.~(\ref{gere}) into Eq.~(\ref{thetaijx}), we obtain
\be
\theta^{ij} (x)=\frac{\mu}{4\pi G_4(\phi_0) D} 
\left( v^i v^j-\tilde{G} m 
\frac{r^i r^j}{r^3} \right)\,,
\label{theij}
\ee
where $v^i \equiv \dot{r}^i=\dot{x}_A^i-\dot{x}_B^i$. 
The relative velocity and displacement between two 
point-like particles affect the value of $\theta^{ij}$ 
at the observed point $x$.

Let us consider a relative circular orbit around the center of mass. 
{}From Eq.~(\ref{gere}), the Newtonian equation along the radial 
direction is given by 
\be
\mu \frac{v^2}{r}=
\frac{\tilde{G} m_A m_B}{r^2}\,,
\label{gere2}
\ee
and hence $v^2=\tilde{G} m/r$.
We introduce the unit vectors $\hat{v}^i$ and $\hat{r}^i$ such that 
$v^i=v \hat{v}^i$ and $r^i=r \hat{r}^i$.
Then, Eq.~(\ref{theij}) reduces to 
\be
\theta^{ij}(x)=\frac{\tilde{G}\mu m}{4\pi G_4(\phi_0) r D}
\left( \hat{v}^i \hat{v}^j-\hat{r}^i \hat{r}^j \right)\,.
\label{thetaijdef}
\ee
This is the leading-order solution to $\theta^{ij}(x)$
for the quasicircular orbit. 
Note that the positions of particles $A$ and $B$ 
can be expressed as 
\be
x_A^i=\frac{\mu}{m_A} r^i\,,\qquad 
x_B^i=-\frac{\mu}{m_B} r^i\,,
\label{xABr}
\ee
together with their velocities 
$\dot{x}_A^i=\mu v^i/m_A$ and 
$\dot{x}_B^i=-\mu v^i/m_B$.

\section{Gravitational waves from compact binary systems}
\label{scapersec}

\subsection{Solutions to scalar-field perturbations}

Since we derived the quadrupole formula (\ref{thetaijdef}) for 
tensor waves, the next procedure is to obtain solutions to 
the scalar-field perturbation $\varphi$. 
For this purpose, we perform the PN expansion 
up to quadrupole order for the scalar-field perturbation equation. 
We first express the derivative term $\square \phi$ by 
using the d'Alembertian $\square_{\rm M}$ 
in Minkowski spacetime as
\be
\square \phi=\left( 1+\frac{1}{2} \theta+g_4 \frac{\varphi}
{\Mpl} \right) \square_{\rm M} \phi
-\theta^{\mu \nu} \partial_{\mu} \partial_{\nu} \varphi
-\frac{g_4}{\Mpl}\partial_{\mu}\varphi\partial^{\mu}\varphi
+{\cal O}(\varepsilon^3)\,,
\ee
where $\theta$ is a trace of the metric tensor $\theta_{\mu \nu}$. 
Up to second order in perturbations $\varphi$ and $\theta$, 
the scalar-field Eq.~(\ref{fieldeq}) is given by 
\ba
\left( \square_{\rm M}-m_s^2 \right) \vp
&=& 
-\frac{1}{\zeta_0} \left( 1-\frac{1}{2}\theta
-g_4 \frac{\varphi}{\Mpl}-\frac{\zeta_1}{\zeta_0} 
\varphi \right) \left( T_{,\phi}-\frac{G_{4,\phi}}{2G_4} 
T \right) \nonumber\\
& & +{\cal O} \left( \varphi^2, \partial_{\mu} \varphi 
\partial^{\mu} \varphi, (\square_{\rm M} \varphi)^2, 
\partial^{\mu}\partial^{\nu} \varphi
\partial_{\mu}\partial_{\nu} \varphi, 
\theta \varphi, 
\theta^{\mu \nu} \partial_{\mu} \partial_{\nu} \varphi \right)\,,
\label{vpeqf}
\ea
where $\zeta_0$ is defined in Eq.~(\ref{ze0def}), and 
\be
\zeta_1 \equiv G_{2,\phi X}-2G_{3,\phi \phi}
+\frac{6 G_{4,\phi} G_{4,\phi \phi}}{G_4}
-\frac{3G_{4,\phi}^3}{(G_4)^2} \biggr|_{\phi=\phi_0}\,.
\ee
We perform a PN expansion of the source term 
corresponding to the first term on the right hand-side 
of Eq.~(\ref{vpeqf}).
In the expression of the trace $T$ given by Eq.~(\ref{trace}), 
we pick up terms up to the orders of $U$, $U_s$, 
and $v_I^2$. We also exploit the expansions 
\ba
& &
\frac{1}{\sqrt{-g}}=1-\frac12 h=1-\frac{1}{8\pi} 
\left[ \frac{U}{G_4(\phi_0)}-\frac{2g_4 U_s}{\zeta_0 \Mpl^2} \right]\,,\\
& &
\frac{1}{u_I^0}=1-\frac12 h_{00}-\frac12 v_I^2=
1-\frac{1}{16\pi} \left[ \frac{U}{G_4(\phi_0)}
+\frac{g_4 U_s}{\zeta_0 \Mpl^2} \right]-\frac12 v_I^2\,,
\ea
as well as Eqs.~(\ref{mAexpan}) and (\ref{G4expan}).
Then, it follows that 
\be
T_{,\phi}-\frac{G_{4,\phi}}{2G_4} T
=
-\sum_{I=A,B} \frac{m_I(\phi_0)}{\Mpl} \left[ 
\hat{\alpha}_I \left( 
1-\frac{3U}{16 \pi G_4(\phi_0)}-\frac12 v_I^2 \right)
+\frac{U_s}{16 \pi \zeta_0 \Mpl^2} 
\left( 2\hat{\alpha}_I^2 +4g_4 \hat{\alpha}_I+2\beta_I
-\gamma_4 \right) 
\right] \delta^{(3)} ({\bm x}-{\bm x}_A(t)).
\label{Tphicon}
\ee
Terms in the second line of Eq.~(\ref{vpeqf}) 
are at most of order $U^2$, $U_s^2$, $UU_s$. 
Since they are higher than quadrupole order in the 
PN expansion, we neglect them in the following discussion. 
In the presence of a cubic derivative interaction $G_3(X) \square \phi$, 
nonlinear terms like $(\square_{\rm M} \varphi)^2$ and 
$\partial^{\mu}\partial^{\nu} \varphi \partial_{\mu}\partial_{\nu} \varphi$
can dominate over the left hand-side of Eq.~(\ref{vpeqf}) within a 
Vainshtein radius $r_V$ \cite{Vainshtein:1972sx,Burrage:2010rs,DeFelice:2011th,Kimura:2011dc,Koyama:2013paa}. 
In the following we assume that $r_V$ is at most of order the 
radius $r_s$ of the star ($r_V \lesssim r_s$), so that the 
PN expansion given below is valid outside the source. 
In other words, our analysis loses its validity for $r_V \gg r_s$ 
due to the dominance of nonlinear derivative terms in the scalar-field 
equation inside the Vainshtein radius.

On using Eq.~(\ref{Tphicon}), the scalar-wave Eq.~(\ref{vpeqf}) 
up to quadrupole order can be expressed as 
\be
\left( \square_{\rm M}-m_s^2 \right) \vp
=-16 \pi S\,,
\label{phieS}
\ee
where the source term is 
\ba
S &=&
-\frac{1}{16 \pi \zeta_0 \Mpl} \sum_{I=A,B} m_I(\phi_0) 
\delta^{(3)} ({\bm x}-{\bm x}_I(t))
\biggl[ \hat{\alpha}_I
\left( 1-\frac{U}{16 \pi G_{4}(\phi_0)}-\frac{1}{2}v_I^2 
\right) \nonumber\\
& &+\frac{U_s}{16 \pi \zeta_0 \Mpl^2} 
\left( 2\hat{\alpha}_I^2 +2g_4 \hat{\alpha}_I+2\beta_I
-\gamma_4-2\hat{\alpha}_I \frac{\Mpl \zeta_1}{\zeta_0} 
\right) \biggr]\,.
\ea
At spatial point ${\bm x}={\bm x}_A$ of the source $A$,  
we have $U({\bm x}_A)=m_B/r$ and $U_s({\bm x}_A)=
-2\hat{\alpha}_B m_B e^{-m_s r}/r$. 
Similarly, at ${\bm x}={\bm x}_B$, 
$U({\bm x}_B)=m_A/r$ and $U_s({\bm x}_B)
=-2\hat{\alpha}_A m_A e^{-m_s r}/r$.
The solution to Eq.~(\ref{phieS}) measured 
by an observer at the position vector  
${\bm D}=D{\bm n}$ and time $t$ is expressed by 
the sum of a ``massless'' solution $\varphi_B$ and 
``massive'' solution $\varphi_m$ \cite{Alsing:2011er,Liu:2020moh}, 
such that 
\be
\varphi=\varphi_B+\varphi_m\,,
\ee
where 
\ba
\varphi_B (t, {\bm D})
&=&4 \int {\rm d}^3 {\bm x}' {\rm d}t'
\frac{S(t',{\bm x}')}{|{\bm D}-{\bm x}'|}
\delta (t-t'-|{\bm D}-{\bm x}'|)\,,\\
\varphi_m (t, {\bm D})
&=&-4 \int {\rm d}^3 {\bm x}' {\rm d}t'
\frac{m_s S(t',{\bm x}') J_1 
(m_s \sqrt{(t-t')^2-|{\bm D}-{\bm x}'|^2})}{\sqrt{(t-t')^2-|{\bm D}-{\bm x}'|^2}}
\Theta (t-t'-|{\bm D}-{\bm x}'|)\,,
\ea
where $J_1$ is a Bessel function of the first kind, and 
$\Theta$ is a Heaviside function.
Far away from the source ($D \gg |{\bm x}'|$), we exploit the 
approximation $|{\bm D}-{\bm x}'| =D-{\bm x}' \cdot {\bm n}$ 
and replace the $t'$ dependence in $S$ with 
$t'=t-D+{\bm x}' \cdot {\bm n}$. 
Performing multipole expansions for the time-dependent 
part of $S$, it follows that 
\ba
\varphi_B (t, {\bm D})
&=& \frac{4}{D} \sum_{\ell=0}^{\infty}
\frac{1}{\ell!} \frac{\partial^{\ell}}{\partial t^{\ell}}
\int {\rm d}^3 {\bm x}' S(t-D, {\bm x}') 
({\bm x}' \cdot {\bm n})^{\ell}\,,
\label{vB1} \\
\varphi_m (t, {\bm D})
&=&-\frac{4}{D} \sum_{\ell=0}^{\infty}
\frac{1}{\ell!} \frac{\partial^{\ell}}{\partial t^{\ell}}
\int {\rm d}^3 {\bm x}' 
({\bm x}' \cdot {\bm n})^{\ell}
\int_0^{\infty} {\rm d}z
\frac{S(t-Du,{\bm x}') J_1(z)}{u^{\ell+1}}\,,
\label{vB2}
\ea
where 
\be
u \equiv \sqrt{1+\frac{z^2}{m_s^2D^2}}\,,\qquad
z \equiv m_s \sqrt{(t-t')^2-|{\bm D}-{\bm x}'|^2}\,.
\ee
We consider a quasicircular orbit of the binary system given by 
the point-like particle equations of motion (\ref{geA}) with Eq.~(\ref{xABr}).
We pick up the contributions up to quadrupole ($\ell=2$) terms 
in Eqs.~(\ref{vB1}) and (\ref{vB2}).
For the dipole and quadrupole contributions, 
we use the following relations
\ba
& & \sum_{I=A,B} m_{I}(\phi_0) \hat{\alpha}_I 
\frac{\partial}{\partial t}
({\bm x}_I \cdot {\bm n})
=\mu \left( \hat{\alpha}_A-\hat{\alpha}_B \right) 
{\bm v} \cdot {\bm n}\,,\\
& &  \sum_{I=A,B} m_{I}(\phi_0) \hat{\alpha}_I
\frac{1}{2!} \frac{\partial^2}{\partial t^2}
({\bm x}_I \cdot {\bm n})^2
=-\frac12 \mu \Gamma ({\bm v}\cdot {\bm n})^2
+\frac{\tilde{G}\mu m}{2r^3}\Gamma 
 ({\bm r}\cdot {\bm n})^2 \,,
\ea
where ${\bm v}=\dot{{\bm x}}_A-\dot{{\bm x}}_B$, $\mu$ and 
$m$ are defined in Eqs.~(\ref{mudefi}) and (\ref{massAB}) 
respectively, and 
\be
\Gamma \equiv -2\frac{m_B \hat{\alpha}_A
+m_A \hat{\alpha}_B}{m}\,.
\ee
There are time-independent contributions to $\varphi_B$ and 
$\varphi_m$ (i.e., those without containing the dependence of 
${\bm r}$ and ${\bm v}$) irrelevant to the gravitational radiation power. 
Dropping such terms,  
we obtain the following solution far away from the source
\ba
\varphi_B
&=& \frac{\mu}{4\pi \zeta_0 \Mpl D}
\biggl\{ \frac{\hat{\alpha}_A+\hat{\alpha}_B}{16 \pi G_4(\phi_0)}\frac{m}{r}
-\frac14 \Gamma v^2-\frac{{\cal F}_s}{16\pi \zeta_0 \Mpl^2}
\frac{m e^{-m_sr}}{r} \nonumber \\
& &
\qquad \qquad \qquad 
 -\left( \hat{\alpha}_A-\hat{\alpha}_B \right){\bm v} \cdot {\bm n}
+\frac12 \Gamma ({\bm v}\cdot {\bm n})^2
-\frac{\Gamma}{2}\frac{\tilde{G}m}{r^3}
 ({\bm r}\cdot {\bm n})^2
\biggr\} \biggr|_{t-D}\,,\label{phiBso} \\
\varphi_m 
&=&-\frac{\mu}{4\pi \zeta_0 \Mpl D}
\biggl\{ \frac{\hat{\alpha}_A+\hat{\alpha}_B}{16 \pi G_4(\phi_0)}
I_1 \left[ \frac{m}{r} \right]
-\frac14 \Gamma I_1 \left[ v^2\right]
-\frac{{\cal F}_s}{16\pi \zeta_0 \Mpl^2}
I_1 \left[ \frac{m e^{-m_sr}}{r}\right] \nonumber \\
& &
\qquad \qquad \qquad 
-\left( \hat{\alpha}_A-\hat{\alpha}_B \right)
I_2 \left[ {\bm v} \cdot {\bm n} \right]
+\frac12 \Gamma I_3 \left[ ({\bm v}\cdot {\bm n})^2 \right]
-\frac{\Gamma}{2} I_3 \biggl[ \frac{\tilde{G}m}{r^3}
 ({\bm r}\cdot {\bm n})^2 \biggr]
\biggr\}\biggr|_{t-Du}\,,\label{phimso}
\ea
where 
\ba
\hspace{-0.7cm}
{\cal F}_s &\equiv&
-2\hat{\alpha}_B
\left( 2\hat{\alpha}_A^2 +2g_4 \hat{\alpha}_A+2\beta_A
-\gamma_4-2\hat{\alpha}_A \frac{\Mpl \zeta_1}{\zeta_0} 
\right)
-2\hat{\alpha}_A
\left( 2\hat{\alpha}_B^2 +2g_4 \hat{\alpha}_B+2\beta_B
-\gamma_4-2\hat{\alpha}_B \frac{\Mpl \zeta_1}{\zeta_0} 
\right)\,,\\
\hspace{-0.7cm}
I_n [f(t)] &\equiv& \int_0^{\infty}{\rm d}z 
\frac{f(t-Du)J_1(z)}{u^n}\,.
\ea
Terms proportional to ${\bm v} \cdot {\bm n}$ correspond to 
the dipole mode, whereas terms proportional to 
$({\bm v} \cdot {\bm n})^2$ and $({\bm r} \cdot {\bm n})^2$ 
represent the quadrupole contributions.
Terms in the first lines of Eqs.~(\ref{phiBso}) and (\ref{phimso}) correspond 
to the monopole mode. Since we are interested in the wavelike
behavior of scalar-field perturbations, we will drop the monopole 
terms in the discussion below.
We note that $\varphi_B$ and $\varphi_m$ acquire 
the time dependence through the changes of ${\bm r}$ 
and ${\bm v}$ induced by the energy loss of gravitational radiation.
We will discuss this issue in Sec.~\ref{waveformsec}.

\subsection{Solutions to GW fields}

The observed GWs at the detector can be quantified by the 
deviation from a geodesic equation. 
The distance $\xi^{i}$ between freely moving 
test particles is modified by the propagation of GWs.
As long as the test particles move slowly and $\xi^{i}$ 
is smaller than the wavelength of GWs, 
the geodesic deviation equation reduces to 
${\rm d}^2 \xi^i/{\rm d}t^2=-R_{0i0j} \xi^j$ \cite{Maggiore:2007ulw}, 
where $R_{0i0j}$'s are components of the Riemann tensor. 
The GW field ${\bm h}_{ij}$ is defined by 
\be
\D_0^2
{\bm h}_{ij}
=-2R_{0i 0j}\,.
\label{hijeq}
\ee
At linear order in $h_{\mu \nu}$, we have 
$R_{0i 0j}=-(\D_0^2 h_{ij}+\D_i \D_j h_{00})/2$. 
We choose the traceless-transverse (TT) gauge 
\be
\partial^j \theta_{ij}=0\,,\qquad 
\theta=0\,,
\ee
under which $h_{00}=g_4 \varphi/\Mpl$.
Then, Eq.~(\ref{hijeq}) yields
\be
\D_0^2 {\bm h}_{ij}
=\D_0^2 \theta_{ij}^{\rm TT}-\delta_{ij}g_4 \frac{\D_0^2 
\varphi}{\Mpl}+g_4 \frac{\D_i \D_j \varphi}{\Mpl}\,,
\label{hijeq2}
\ee
where ``TT'' represents the TT gauge.
The solution to $\varphi$ without the monopole terms
is expressed as 
\be
\varphi=\varphi_B (t-D, {\bm n})+\varphi_m (t-Du, {\bm n})\,,
\ee
where 
\ba
\varphi_B(t-D,{\bm n})
&=&-\frac{\mu}{4\pi \zeta_0 \Mpl D} 
\biggl\{ \left( \hat{\alpha}_A-\hat{\alpha}_B \right){\bm v} \cdot {\bm n}
-\frac12 \Gamma ({\bm v}\cdot {\bm n})^2
+\frac{\Gamma}{2}\frac{\tilde{G}m}{r^3}
 ({\bm r}\cdot {\bm n})^2
\biggr\}\,,\label{varphiB2} 
\\
\varphi_m(t-Du,{\bm n})
&=& \frac{\mu}{4\pi \zeta_0 \Mpl D} 
\int_0^{\infty}{\rm d}z J_1(z) \psi_m\,,
\label{varphim2} 
\ea
where
\be
\psi_m=\left( \hat{\alpha}_A-\hat{\alpha}_B \right)
\frac{{\bm v} \cdot {\bm n}}{u^2}
-\frac12 \Gamma \frac{({\bm v}\cdot {\bm n})^2}{u^3}
+\frac{\Gamma}{2}\frac{\tilde{G}m}{r^3}
\frac{({\bm r}\cdot {\bm n})^2}{u^3}\biggr|_{t-Du}\,.
\ee
To compute the last term of Eq.~(\ref{hijeq2}), we exploit the 
following properties \cite{Liu:2018sia}
\ba
& &
\D_i \D_j \varphi_B (t-D,{\bm n})=n_i n_j 
\D_0^2 \varphi_B (t-D,{\bm n})\,,\\
& &
\D_i \D_j \psi_m (t-Du,{\bm n})=
\frac{n_i n_j}{u^2} 
\D_0^2 \psi_m (t-Du,{\bm n})\,,
\ea
where $n_i=x_i/D$, and we ignored the next-to-leading order 
contributions to $\D_i \D_j \varphi_B$ arising from the spatial 
derivative of the term proportional to $1/D$ 
in Eq.~(\ref{varphiB2}) (and likewise for $\D_i \D_j \psi_m$).
Then, Eq.~(\ref{hijeq2}) reduces to 
\be
\D_0^2 {\bm h}_{ij}
=\D_0^2 \left[ \theta_{ij}^{\rm TT}-
\left( \delta_{ij}-n_i n_j \right)
g_4 \frac{\varphi}{\Mpl}
+n_i n_j \frac{g_4 \mu}{4\pi \zeta_0 \Mpl^2 D} 
\int_0^{\infty}{\rm d}z J_1(z)
\left( \frac{1}{u^2}-1 \right)\psi_m
\right]\,.
\label{hijeq3}
\ee
In the three dimensional Cartesian coordinate $(x_1, x_2, x_3)$,  
we consider the GWs propagating along the $x_3$ direction, 
in which case $n_{x_1}=n_{x_2}=0$ and $n_{x_3}=1$. 
We express the GW field in the form 
\ba
{\bm h}_{ij}=
\left(
\begin{array}{ccc}
h_{+}+h_{b} & h_{\times} & 0 \\
h_{\times}  & -h_{+}+h_{b} & 0 \\
0 & 0 & h_L \\
\end{array}
\right)\,.
\ea
{}From Eq.~(\ref{hijeq3}), it follows that 
\ba
& &
h_{+}=\theta_{11}^{\rm TT}=-\theta_{22}^{\rm TT}\,,\qquad
h_{\times}=\theta_{12}^{\rm TT}=\theta_{21}^{\rm TT}\,,
\label{hthe}\\
& &
h_b=-g_4 \frac{\varphi}{\Mpl}\,,
\label{hb}\\
& &
h_L=\frac{g_4 \mu}{4\pi \zeta_0 \Mpl^2 D} 
\int_0^{\infty}{\rm d}z J_1(z)
\left( \frac{1}{u^2}-1 \right)\psi_m\,.
\label{hL}
\ea
Besides the TT polarizations $h_{+}$ and $h_{\times}$, 
the presence of a nonminimally coupled scalar 
field ($g_4 \neq 0$) gives rise to a breathing mode 
$h_b$ and a longitudinal mode $h_L$ \cite{Eardley:1973zuo,Maggiore:1999wm}.
The longitudinal mode, which has a polarization along the propagating 
direction of GWs, arises from the nonvanishing mass $m_s$ of the scalar field.

In the Cartesian coordinate system $(x_1,x_2,x_3)$ whose origin O coincides 
with the center of mass of the binary system, 
we choose the unit vector field ${\bm n}$ from O to the observer 
in the $(x_2,x_3)$ plane with an angle $\gamma$ inclined from 
the $x_3$ axis.
The quasicircular motion of the binary system, which is characterized 
by the relative vector ${\bm r}$, is confined on the $(x_1,x_2)$ plane 
with an angle $\Phi$ inclined from the $x_1$ axis. 
Then, we can express the unit vectors ${\bm n}$, $\hat{{\bm r}}$, 
and $\hat{{\bm v}}$ as
\be
{\bm n}=(0, \sin \gamma, \cos \gamma)\,,\qquad 
\hat{{\bm r}}=(\cos \Phi, \sin \Phi, 0)\,,\qquad 
\hat{{\bm v}}=(-\sin \Phi, \cos \Phi, 0)\,,
\ee
with ${\bm r}=r\hat{{\bm r}}$ and ${\bm v}=v\hat{{\bm v}}$. 

{}From Eq.~(\ref{thetaijdef}), the TT components 
of $\theta_{ij}$ for the GWs propagating along the $x_3$ axis are 
$\theta_{x_1 x_1}=-\theta_{x_2 x_2}=-A_{\theta} \cos(2\Phi)$ and 
$\theta_{x_1 x_2}=\theta_{x_2 x_1}=-A_{\theta} \sin (2\Phi)$, where 
$A_{\theta}=\tilde{G}\mu m/[4\pi G_4(\phi_0) r D]$. 
After rotation of the angle $\gamma$,  
the GWs propagating along the direction of 
${\bm n}$ have the components $\tilde{\theta}_{11}=\theta_{x_1 x_1}$, 
$\tilde{\theta}_{12}=\tilde{\theta}_{21}=\theta_{x_1 x_2}\cos \gamma$, and 
$\tilde{\theta}_{22}=\theta_{x_2 x_2} \cos^2 \gamma$. 
The TT components $\theta_{ij}^{\rm TT}$ 
are derived by using a Lambda tensor $\Lambda_{ij,kl}$ \cite{Maggiore:2007ulw}, 
as $\theta_{ij}^{\rm TT}=\Lambda_{ij,kl}\tilde{\theta}_{kl}$. 
Since $\theta_{11}^{\rm TT}=-\theta_{22}^{\rm TT}
=(\tilde{\theta}_{11}-\tilde{\theta}_{22})/2$ and 
$\theta_{12}^{\rm TT}=\theta_{21}^{\rm TT}=
\tilde{\theta}_{12}$, we obtain the following TT components
\ba
h_{+} &=&
-(1+\delta)^{2/3} \frac{4(G_* M_c)^{5/3} \omega^{2/3}}{D}
\frac{1+\cos^2 \gamma}{2} \cos(2\Phi)\,,\label{hp}\\
h_{\times} 
&=&-(1+\delta)^{2/3} \frac{4(G_* M_c)^{5/3} \omega^{2/3}}{D}
\cos \gamma \sin(2\Phi)\,,\label{ht}
\ea
where $\omega=v/r$ is an orbital frequency, 
$M_c=\mu^{3/5}m^{2/5}$ is a 
chirp mass, and 
\be
G_* \equiv \frac{1}{16 \pi G_4(\phi_0)}\,,\qquad
\delta \equiv
4\kappa_4 \hat{\alpha}_A \hat{\alpha}_B
\left( 1+m_s r \right) e^{-m_s r}\,,\qquad 
\kappa_4 \equiv
\frac{G_4(\phi_0)}{\zeta_0 \Mpl^2}\,,
\label{Gstar}
\ee
with $\tilde{G}=G_* (1+\delta)$.
{}From Eqs.~(\ref{varphiB2}), (\ref{varphim2}), (\ref{hb}), and 
(\ref{hL}), the breathing and longitudinal modes of 
${\bm h}_{ij}$ are expressed as
\ba
h_b &=& \frac{g_4 \mu}{4\pi \zeta_0 \Mpl^2 D} 
\biggl\{ (\hat{\alpha}_A-\hat{\alpha}_B) v \sin \gamma \cos \Phi
-\frac{1}{2}\Gamma v^2 \sin^2 \gamma \cos(2\Phi) 
\nonumber \\
& &-\int_{0}^{\infty}{\rm d}z\,J_1(z) 
\left[ \frac{1}{u^2} (\hat{\alpha}_A-\hat{\alpha}_B) v \sin \gamma \cos \Phi
-\frac{1}{2u^3}\Gamma v^2 \sin^2 \gamma \cos(2\Phi) 
\right]\biggl\} \,,
\label{hbf}\\
h_L &=& \frac{g_4 \mu}{4\pi \zeta_0 \Mpl^2 D} 
\int_{0}^{\infty}{\rm d}z\,J_1(z) \left( \frac{1}{u^2}-1 
\right)
\left[ \frac{1}{u^2} (\hat{\alpha}_A-\hat{\alpha}_B) v \sin \gamma \cos \Phi
-\frac{1}{2u^3}\Gamma v^2 \sin^2 \gamma \cos(2\Phi) 
\right]\,,
\label{hLf}
\ea
where we used the relation $v^2=\tilde{G}m/r$. 
Performing the integrals in Eqs.~(\ref{hbf}) and (\ref{hLf}) with the limit 
$D \to \infty$, one can show that both $h_b$ and $h_L$ are nonvanishing 
only for high frequency modes with 
$\omega \gtrsim m_s$ \cite{Liu:2018sia,Liu:2020moh}.
The longitudinal mode $h_L$ is about $m_s^2/\omega^2$ times 
as large as the breathing mode $h_b$. 
We note that both $h_b$ and $h_L$ vanish in the limit
$\hat{\alpha}_I \to 0$ (where $I=A,B$).
For the scalar charge in the range $|\hat{\alpha}_I| \ll 1$, 
$h_b$ and $h_L$ are generally suppressed in comparison to 
$h_{+}$ and $h_{\times}$.

\section{Fourier-transformed gravitational waveforms}
\label{waveformsec}

The inspiraling compact binary loses the energy through gravitational 
radiation. This leads to the time variation of the orbital frequency $\omega$. 
In this section, we derive gravitational waveforms of TT polarizations 
in the frequency domain under a stationary phase approximation. 

\subsection{Time variation of orbital frequency}

In Refs.~\cite{Hou:2017cjy,AbhishekChowdhuri:2022ora}, the effective GW stress-energy tensor 
$t_{\mu \nu}$ in Horndeski theories was derived under a 
short-wavelength approximation. 
This is based on the approximation that the wavelength of GWs  
is much smaller than a typical background curvature 
scale \cite{Isaacson:1968zza}. 
In the TT gauge, the explicit form of $t_{\mu \nu}$ is given by 
\be
t_{\mu \nu}=\bigg\langle \frac{1}{2} G_4(\phi_0) 
\partial_{\mu} \theta_{\alpha \beta}^{\rm TT} 
\partial_{\nu} \theta^{\alpha \beta}_{\rm TT} 
+\zeta_0 \partial_{\mu} \varphi \partial_{\nu} \varphi
+m_s^2 G_{4,\phi}(\phi_0) \varphi\,\theta_{\mu \nu}^{\rm TT}
\bigg\rangle\,,
\label{taumn}
\ee
where the symbol $\langle \cdots \rangle$ represents the 
time average over an orbital period.
The conservation of $t^{\mu \nu}$ inside a volume $V$
implies that $\int_V {\rm d}^3 x\,(\partial_0 t^{00}+\partial_i t^{0i})=0$.
Thus, the time derivative of the GW energy $E_{\rm GW}=\int_V{\rm d}^3 x\,t^{00}$ is
\be
\frac{{\rm d}E_{\rm GW}}{{\rm d}t}=
-\int_V {\rm d}^3 x\,\partial_i t^{0i}
=-\int_S {\rm d}A\,\hat{N}_i t^{0i}\,,
\ee
where $\hat{N}_i$ is an outer normal to the surface, and 
the last term represents a surface integral.
Taking the surface of a sphere with the radius $D$
and using the property 
$\partial_0 \theta^{\rm TT}_{\alpha \beta}(t-D)=
-\partial_D \theta^{\rm TT}_{\alpha \beta}(t-D)$ 
with Eq.~(\ref{hthe}), 
it follows that 
\be
\frac{{\rm d}E_{\rm GW}}{{\rm d}t}
=-\int_S {\rm d}A\,t^{0D}
=-\int {\rm d}\Omega\,D^2 \left[ 
G_4(\phi_0) \langle \dot{h}_{+}^2+\dot{h}_{\times}^2 
\rangle-\zeta_0 \langle\partial_0 \varphi \partial_{D}\varphi 
\rangle\right]\,,
\label{dEGW}
\ee
where $\Omega$ is the solid angle element. 
On using Eqs.~(\ref{hp}) and (\ref{ht}) with 
$\Phi=\omega (t-D)$, we obtain
\be
-\int {\rm d}\Omega\,D^2 
G_4(\phi_0) \langle \dot{h}_{+}^2+\dot{h}_{\times}^2 
\rangle
=-\frac{512}{5}\pi G_4(\phi_0) (1+\delta)^{4/3} 
(G_* M_c \omega)^{10/3}\,.
\ee
The scalar-field perturbation $\varphi$ is the sum of 
$\varphi_B$ and $\varphi_m$ given by Eqs.~(\ref{varphiB2}) 
and (\ref{varphim2}), respectively. 
{}From Eqs.~(\ref{gere}) and (\ref{gere2}) as well as 
the relation ${\rm d}(Du)/{\rm d}D=1/u$, we obtain 
\ba
\partial_0 \varphi &=& \frac{\tilde{G} \mu m}{4\pi \zeta_0 \Mpl D}
\left\{ (\hat{\alpha}_A-\hat{\alpha}_B) \left( \frac{{\bm r} \cdot {\bm n}}{r^3} 
-I_2 \left[ \frac{{\bm r} \cdot {\bm n}}{r^3} \right] \right)
-2\Gamma \left( \frac{({\bm r} \cdot {\bm n})
({\bm v} \cdot {\bm n})}{r^3}-I_3 \left[ 
\frac{({\bm r} \cdot {\bm n})
({\bm v} \cdot {\bm n})}{r^3}
\right] \right)
\right\}\,,\\
\partial_D \varphi &=& -\frac{\tilde{G} \mu m}{4\pi \zeta_0 \Mpl D}
\left\{ (\hat{\alpha}_A-\hat{\alpha}_B) \left( \frac{{\bm r} \cdot {\bm n}}{r^3} 
-I_3 \left[ \frac{{\bm r} \cdot {\bm n}}{r^3} \right] \right)
-2\Gamma \left( \frac{({\bm r} \cdot {\bm n})
({\bm v} \cdot {\bm n})}{r^3}-I_4 \left[ 
\frac{({\bm r} \cdot {\bm n})
({\bm v} \cdot {\bm n})}{r^3}
\right] \right)
\right\}\,,
\ea
where ${\bm r} \cdot {\bm n}=r \sin \gamma \sin \Phi$ and 
${\bm v} \cdot {\bm n}=v \sin \gamma \cos \Phi$. 
For the quantities like ${\bm r} \cdot {\bm n}/r^3$, the 
angle $\Phi$ has the dependence $\Phi=\omega (t-D)$, 
while, for the quantities like $I_2[{\bm r} \cdot {\bm n}/r^3]$, 
$\Phi=\omega (t-Du)$. 
Taking the time average of $\partial_0 \varphi \partial_D \varphi$ 
over the orbital period, it follows that 
\ba
\langle\partial_0 \varphi \partial_{D}\varphi \rangle
&=& 
-\left( \frac{\tilde{G} \mu m}{4\pi \zeta_0 \Mpl D} \right)^2 
\biggl\langle (\hat{\alpha}_A-\hat{\alpha}_B)^2 
\left( \frac{{\bm r} \cdot {\bm n}}{r^3} 
-I_2 \left[ \frac{{\bm r} \cdot {\bm n}}{r^3} \right] \right)
 \left( \frac{{\bm r} \cdot {\bm n}}{r^3} 
-I_3 \left[ \frac{{\bm r} \cdot {\bm n}}{r^3} \right] \right) 
\nonumber \\
& &+4\Gamma^2\left( \frac{({\bm r} \cdot {\bm n})
({\bm v} \cdot {\bm n})}{r^3}-I_3 \left[ 
\frac{({\bm r} \cdot {\bm n})
({\bm v} \cdot {\bm n})}{r^3}
\right] \right)
\left( \frac{({\bm r} \cdot {\bm n})
({\bm v} \cdot {\bm n})}{r^3}-I_4 \left[ 
\frac{({\bm r} \cdot {\bm n})
({\bm v} \cdot {\bm n})}{r^3}
\right] \right) \biggr\rangle\,.
\label{provarphi}
\ea
For the computation on the right hand-side of Eq.~(\ref{provarphi}), 
we introduce the following integrals
\ba
& &
C_n=\int_0^{\infty}{\rm d}z\,\cos (\omega D u) 
\frac{J_1(z)}{u^n}\,,\qquad 
S_n=\int_0^{\infty}{\rm d}z\,\sin (\omega D u) 
\frac{J_1(z)}{u^n}\,,\\
& &
\tilde{C}_n=\int_0^{\infty}{\rm d}z\,\cos (2\omega D u) 
\frac{J_1(z)}{u^n}\,,\qquad 
\tilde{S}_n=\int_0^{\infty}{\rm d}z\,\sin (2\omega D u) 
\frac{J_1(z)}{u^n}\,.
\ea
Then, the last integral in Eq.~(\ref{dEGW}) reads 
\ba
\int {\rm d}\Omega\,D^2 \zeta_0 
\langle\partial_0 \varphi \partial_{D}\varphi \rangle
&=&-\frac{(\tilde{G} \mu m)^2}{12\pi \zeta_0 \Mpl^2 r^4} 
\biggl[ (\hat{\alpha}_A-\hat{\alpha}_B)^2 \left\{ 1-\cos(\omega D) 
(C_2+C_3)-\sin (\omega D) (S_2+S_3)+C_2 C_3+S_2 S_3 
\right\} \nonumber \\
& &+\frac{4}{5}\Gamma^2 v^2 \left\{ 1-\cos(2\omega D) 
(\tilde{C}_3+\tilde{C}_4)-\sin (2\omega D) 
(\tilde{S}_3+\tilde{S}_4)+\tilde{C}_3 \tilde{C}_4 
+\tilde{S}_3 \tilde{S}_4  \right\} \biggr]\,.
\ea
In the large-distance limit $D \to \infty$, the asymptotic forms 
of $C_n$ and $S_n$ are given, respectively, by 
\ba
C_n &\simeq& \cos (\omega D)-\left( 1-\frac{m_s^2}{\omega^2} 
\right)^{(n-1)/2} \cos \left( D \sqrt{\omega^2-m_s^2} 
\right) \Theta (\omega-m_s)\,,
\label{Cna} \\
S_n &\simeq& \sin (\omega D)-\left( 1-\frac{m_s^2}{\omega^2} 
\right)^{(n-1)/2} \sin \left( D \sqrt{\omega^2-m_s^2} 
\right) \Theta (\omega-m_s)\,.
\label{Dna} 
\ea
As for $\tilde{C}_n$ and $\tilde{S}_n$ in the limit 
$D \to \infty$, we only need to change $\omega$ 
in Eqs.~(\ref{Cna}) and (\ref{Dna}) to $2\omega$, respectively. 
Then, at large $D$, the energy loss of GWs induced by the stress-energy 
tensor $t_{\mu \nu}$ yields
\ba
\frac{{\rm d}E_{\rm GW}}{{\rm d}t}
&=& 
-\frac{512}{5}\pi G_4(\phi_0) (1+\delta)^{4/3} 
(G_* M_c \omega)^{10/3} \nonumber \\
&&-\frac{(\tilde{G} \mu m)^2}{12\pi \zeta_0 \Mpl^2 r^4} 
\left[ (\hat{\alpha}_A-\hat{\alpha}_B)^2
\left( 1-\frac{m_s^2}{\omega^2} \right)^{3/2} 
\Theta (\omega-m_s)+
\frac{4}{5} \Gamma^2 v^2 
\left( 1-\frac{m_s^2}{4\omega^2} \right)^{5/2} 
\Theta (2\omega-m_s) \right]\,,
\label{dEGWf}
\ea
where the terms on the second line arise from scalar radiation. 
For the frequency in the range $\omega<m_s/2$
there is no scalar radiation, but, for $\omega>m_s$, 
the two terms in the square bracket of (\ref{dEGWf}) 
are nonvanishing. 

The energy $E$ associated with the binary system is given by 
\be
E=\frac{1}{2}\mu v^2-\tilde{G}\frac{\mu m}{r}
=-\frac{\tilde{G} \mu m}{2r}
=-\frac{1}{2} \mu \left( \tilde{G} m \omega \right)^{2/3}\,.
\ee
The orbital frequency $\omega$ changes in time due to the 
decrease of $E$ induced by the energy loss $E_{\rm GW}$. 
Since ${\rm d}E/{\rm d}t={\rm d}E_{\rm GW}/{\rm d}t$, we obtain 
the time variation of $\omega$, as 
\ba
\dot{\omega}
&=& 
\frac{96}{5}(G_* M_c)^{5/3} \omega^{11/3} 
\biggl[
\left( 1+\delta \right)^{2/3} \nonumber \\
&&+\frac{5}{24} \kappa_4 
\biggl\{ (\hat{\alpha}_A-\hat{\alpha}_B)^2
\left( 1-\frac{m_s^2}{\omega^2} \right)^{3/2}
\frac{\Theta (\omega-m_s)}{(G_* m \omega)^{2/3}} 
+
\frac{4}{5} \Gamma^{2} (1+\delta)^{2/3}
\left( 1-\frac{m_s^2}{4\omega^2} \right)^{5/2} 
\Theta (2\omega-m_s) \biggr\} \biggr]\,,
\label{dotomega}
\ea
where we used the relation 
\be
\frac{\mu \omega^3}{4\pi \zeta_0 \Mpl^2}
=4\kappa_4 \frac{(G_* M_c)^{5/3} \omega^{11/3}}
{(G_* m \omega)^{2/3}}\,.
\ee
We recall that  $G_*$, $\delta$, and $\kappa_4$ 
are defined in Eq.~(\ref{Gstar}).

\subsection{Gravitational waveforms}
\label{GWwavesec}

When we confront the gravitational waveform with observations, 
it is common to perform a Fourier transformation of $h_+$, 
$h_{\times}$, $h_b$, and $h_L$ with a frequency $f$.
Since the amplitudes of $h_+$ and $h_{\times}$ are typically 
much larger than those of $h_b$ and $h_L$ \cite{Liu:2018sia,Liu:2020moh}, 
we will estimate the deviation from GR for the two polarizations $h_+$ and $h_{\times}$ in Fourier space. Let us perform the Fourier transformation 
\be
\tilde{h}_{\lambda} (f)=\int {\rm d}t\,h_{\lambda}(t) 
e^{i \cdot 2 \pi f t}\,,
\ee
where $\lambda=+, \times$. 
For the $\lambda=+$ mode, using Eq.~(\ref{hp}) with 
$\Phi=\omega (t-D)$ leads to  
\be
\tilde{h}_{+} (f)=
-(1+\delta)^{2/3} \frac{(G_* M_c)^{5/3}}{D}
(1+\cos^2 \gamma)\,
e^{i \cdot 2 \pi f D}
\int {\rm d}t\, \omega(t)^{2/3} 
\left[ e^{i(2\Phi(t)+2 \pi ft)}
+e^{i(-2\Phi(t)+2 \pi ft)} \right]\,.
\label{hp2}
\ee
There is a stationary phase point for 
the second term in the square bracket of Eq.~(\ref{hp2}), 
such that 
\be
\frac{{\rm d}}{{\rm d}t} \left[ -2\Phi(t)+ 
2\pi ft \right]\biggr|_{t=t_*}=0 \qquad 
\to \qquad \omega(t_*)=\pi f\,.
\ee
Since the first term in the square bracket of Eq.~(\ref{hp2}) 
exhibits fast oscillations, we ignore such a contribution 
in the following discussion. 
Expanding $\Phi(t)$ around $t=t_*$ as 
$\Phi(t)=\Phi(t_*)+\pi f (t-t_*)+\dot{\omega}(t_*)(t-t_*)^2/2
+{\cal O}((t-t_*)^3)$, 
it follows that 
\be
\tilde{h}_{+} (f)=
-(1+\delta)^{2/3} \frac{(G_* M_c)^{5/3}}{D}
(1+\cos^2 \gamma)\,
e^{i [2\pi fD-2\Phi(t_*)+2\pi f t_*]}
\int {\rm d}t\, \omega(t)^{2/3} 
e^{-i \dot{\omega}(t_*) (t-t_*)^2}\,.
\label{hp3}
\ee
Since $\int {\rm d}t\, \omega(t)^{2/3} 
e^{-i \dot{\omega}(t_*) (t-t_*)^2} \simeq \omega(t_*)^{2/3} 
\sqrt{\pi/\dot{\omega}(t_*)}\,e^{-i \pi/4}$, 
we obtain 
\be
\tilde{h}_{+} (f)=
-(1+\delta)^{2/3} \frac{(G_* M_c)^{5/3}}{D}
(1+\cos^2 \gamma)\,\omega(t_*)^{2/3} 
\sqrt{\frac{\pi}{\dot{\omega}(t_*)}}
e^{i \Psi_+}\,,
\label{hp4}
\ee
where 
\be
\Psi_{+}=2\pi f t_*-2\Phi(t_*)+2\pi fD-\frac{\pi}{4}\,.
\label{Psip}
\ee
Similarly, the Fourier-transformed mode of 
$h_{\times}(f)$ is given by 
\be
\tilde{h}_{\times} (f)=
-2(1+\delta)^{2/3} \frac{(G_* M_c)^{5/3}}{D}
(\cos \gamma)\,\omega(t_*)^{2/3} 
\sqrt{\frac{\pi}{\dot{\omega}(t_*)}}
e^{i \Psi_\times}\,,
\label{hm4}
\ee
where 
\be
\Psi_{\times}=\Psi_{+}+\frac{\pi}{2}\,.
\ee
The orbital frequency $\omega$ increases according to Eq.~(\ref{dotomega}). 
At a critical time $t_c$, $\omega$ grows sufficiently large, 
such that $\omega(t_c) \to \infty$.
Then, the time $t_*$ can be expressed as 
\be
2\pi f t_*-2\Phi(t_*)=2\pi f t_c-2\Phi_c
+\int_{\infty}^{\pi f} {\rm d} \omega \frac{2\pi f-2\omega}{\dot{\omega}}\,,
\ee
where $\Phi_c=\Phi(t_c)$. 
Substituting this relation into Eq.~(\ref{Psip}), it follows that 
\be
\Psi_{+}=2\pi f \left( D+t_c \right)-2\Phi_c-\frac{\pi}{4}
+\int_{\infty}^{\pi f} {\rm d} \omega \frac{2\pi f-2\omega}{\dot{\omega}}\,,
\ee
where the last integral should be performed after the 
substitution of Eq.~(\ref{dotomega}). 

It is important to recognize that terms in the second line 
of Eq.~(\ref{dotomega}) vanish for $\omega<m_s/2$, 
whereas this is not the case for $\omega>m_s/2$. 
Moreover, $\dot{\omega}$ contains the term $\delta$, 
whose behavior is different dependent on whether the 
mass is in the range $m_s r \ll 1$ or $m_s r \gg 1$. 
Using the quasicircular equation of motion $v^2=\tilde{G}m/r$ with $v=r \omega$ 
and $\omega=2\pi f$, the relative distance between the binary system 
is given by 
\be
r=\left( \frac{c^2 r_g}{8 \pi^2 f^2} \right)^{1/3}
=(1.7 \times 10^5\,{\rm m})
\left( \frac{f}{50\,{\rm Hz}} \right)^{-2/3}
\left( \frac{r_g}{10^4\,{\rm m}} \right)^{1/3}\,
\,,
\ee
where $r_g=2\tilde{G}m/c^2$ and we restored the speed of 
light $c$ for the numerical computation. 
The critical scalar mass $\tilde{m}_s$ corresponding to 
$\tilde{m}_s r=1$ can be estimated as 
\be
\tilde{m}_s=\frac{1}{r} \simeq 10^{-12}~{\rm eV}
\left( \frac{f}{50\,{\rm Hz}} \right)^{2/3}
\left( \frac{r_g}{10^4\,{\rm m}} \right)^{-1/3}\,,
\ee
so that $\tilde{m}_s \simeq 10^{-12}$~eV
for the typical frequency $f=50$~Hz during the inspiral phase 
with $r_g=10^4$~m. 
In the heavy mass range $m_s \gg \tilde{m}_s$, 
we have $\delta \simeq 0$ due to the suppression arising from 
the exponential factor $e^{-m_s r}$. 
For $m_s \ll \tilde{m}_s$, $\delta$ has a constant value 
\be
\delta_0=4\kappa_4 \hat{\alpha}_A \hat{\alpha}_B\,.
\ee
We note that the frequency $f=50$~Hz corresponds to 
the order $\omega=2\pi f \simeq 10^{-13}$~eV.
Provided the mass $m_s$ is in the range 
$m_s<\omega \simeq 10^{-13}$~eV, we have $m_s r < 0.1$ 
and hence $\delta$ can be approximated as $\delta_0$.
In the following, we will first consider the light mass region  
$m_s \ll \omega$ and then proceed to the discussion in 
the massive limit $m_s \gg \omega$.

\subsubsection{$m_s \ll \omega$}

For $m_s<\omega$, terms in the second line of Eq.~(\ref{dotomega}), 
which correspond to scalar radiation, are nonvanishing. 
In the limit that $m_s \ll \omega$, we have
\be
\dot{\omega}=
\frac{96}{5}(G_* M_c)^{5/3} \omega^{11/3} 
\left[ (1+\delta_0)^{2/3} \left( 1+\frac{\kappa_4}{6} \Gamma^2 
\right)
+\frac{5 \kappa_4 (\hat{\alpha}_A-\hat{\alpha}_B)^2}
{24 (G_* m \omega)^{2/3}} \right]\,.
\label{dome}
\ee
If $\omega$ is not much different from $m_s$, 
there are corrections arising from the terms $(1-m_s^2/\omega^2)^{3/2}$ 
and $[1-m_s^2/(4\omega^2)]^{5/2}$. 
We ignored such higher-order corrections to the right 
hand-side of Eq.~(\ref{dome}). 
Under the conditions
\be
|\delta_0| \ll 1\,,\qquad 
|\kappa_4 \Gamma^2| \ll 1\,,
\label{delkappa}
\ee
Eq.~(\ref{dome}) can be approximated as
\be
\dot{\omega} \simeq 
\frac{96}{5}(G_* M_c)^{5/3} \omega^{11/3} 
\left[ 1+\frac{2}{3} \delta_0+\frac{\kappa_4}{6} \Gamma^2+
\frac{5 \kappa_4 (\hat{\alpha}_A-\hat{\alpha}_B)^2}
{24 (G_* m \omega)^{2/3}} \right]\,,
\label{dome2}
\ee
where $\delta_0$ and $\kappa_4 \Gamma^2$ 
are at most of order $\kappa_4 \hat{\alpha}_I^2$. 
Since $(G_* m \omega)^{2/3} \approx v^2$, 
the last term in the square bracket of Eq.~(\ref{dome2})
is at most of order $\kappa_4 \hat{\alpha}_I^2(c^2/v^2)$, 
where we restored the speed of light $c$. 
Then, under the PN expansion, the leading-order correction 
to $\dot{\omega}$ arising from the modification to gravity 
is the last term in the square bracket of Eq.~(\ref{dome2}). 
As long as the condition 
\be
\epsilon \equiv \frac{5 \kappa_4 (\hat{\alpha}_A-\hat{\alpha}_B)^2}
{24 (G_* m \omega)^{2/3}} \ll 1
\label{epscon}
\ee
is satisfied together with inequalities (\ref{delkappa}), 
we have
$1/\dot{\omega} \simeq (5/96)(G_* M_c)^{-5/3}
\omega^{-11/3}(1-2\delta_0/3-\kappa_4 \Gamma^2/6-\epsilon)$ 
approximately. 
Then, the phase terms are integrated to give 
\ba
\Psi_{+}= 
\Psi_{\times}-\frac{\pi}{2}
= 2\pi f \left( D+t_c \right)-2\Phi_c-\frac{\pi}{4}
+\frac{3}{128} (G_* M_c \pi f)^{-5/3}
\left[ 1-\frac{2}{3}\delta_0-\frac{\kappa_4}{6}\Gamma^2
-\frac{5 \kappa_4 (\hat{\alpha}_A-\hat{\alpha}_B)^2}
{42 (G_* m \pi f)^{2/3}} \right]\,,
\label{Psiba}
\ea
where we ignored corrections higher than 
the orders $\kappa_4 \hat{\alpha}_I^2(c^2/v^2)$ and 
$\kappa_4 \hat{\alpha}_I^2$.
We also obtain 
\ba
\tilde{h}_{+} (f)
&=& -\frac{(G_* M_c)^{5/6}}{D}
(1+\cos^2 \gamma)\,\sqrt{\frac{5\pi}{96}}\,
(\pi f)^{-7/6} 
\left[ 1+\frac{1}{3} \delta_0-\frac{\kappa_4}{12} \Gamma^2-\frac{5 \kappa_4 (\hat{\alpha}_A-\hat{\alpha}_B)^2}{
48(G_* m \pi f)^{2/3}} 
\right] e^{i \Psi_+} \,,\label{hpf}\\
\tilde{h}_{\times}(f)
&=& -2\frac{(G_* M_c)^{5/6}}{D} (\cos \gamma)
\,\sqrt{\frac{5\pi}{96}}\,(\pi f)^{-7/6} 
 \left[ 1+\frac{1}{3} \delta_0-\frac{\kappa_4}{12} \Gamma^2
 -\frac{5 \kappa_4 (\hat{\alpha}_A-\hat{\alpha}_B)^2}
 {48(G_* m \pi f)^{2/3}} 
\right] e^{i \Psi_\times}\,.
\label{htf}
\ea
If we take higher-order PN corrections into account, 
they appear as the form $1+{\cal O}(v^2/c^2)+\cdots$ 
in the square brackets of Eqs.~(\ref{Psiba})-(\ref{htf}). 
Unlike the $\delta_0$ and $\kappa_4 \Gamma^2$ terms, 
such PN corrections depend on the frequency $f$.

\subsubsection{$m_s \gg \omega$}

For $\omega<m_s/2$ we have $\Theta (\omega-m_s)=0$ and 
$\Theta (2\omega-m_s)=0$, so there is no scalar radiation in Eq.~(\ref{dotomega}). 
For the orbital frequency $\omega \simeq 10^{-13}$~eV 
with the distance $r \simeq 10^{12}$~eV$^{-1}$, 
we have $\omega r \simeq 0.1$.
Then, in the mass range $m_s \gtrsim 10^2\omega=10^{-11}$~eV, 
we have $\delta \ll 1$ and hence 
$\dot{\omega} \simeq (96/5) (G_* M_c)^{5/3} 
\omega^{11/3}$. For such heavy scalar masses, 
$\Psi_{+}, \Psi_{\times}$ and $\tilde{h}_{+}, \tilde{h}_{\times}$ 
reduce to the values in GR as   
\be
\Psi_{+}^{\rm GR}=\Psi_{\times}^{\rm GR}-\frac{\pi}{2}
=2\pi f \left( D+t_c \right)-2\Phi_c-\frac{\pi}{4}
+\frac{3}{128} (G_* M_c \pi f)^{-5/3}\,,
\ee
and 
\ba
\tilde{h}_{+}^{\rm GR} (f)
&=& -\frac{(G_* M_c)^{5/6}}{D}
(1+\cos^2 \gamma)\,\sqrt{\frac{5\pi}{96}}\,
(\pi f)^{-7/6} e^{i \Psi_+^{\rm GR}}\,,\label{htGR}\\
\tilde{h}_{\times}^{\rm GR} (f)
&=& -2\frac{(G_* M_c)^{5/6}}{D} (\cos \gamma)
\,\sqrt{\frac{5\pi}{96}}\,(\pi f)^{-7/6} e^{i \Psi_\times^{\rm GR}}\,.
\label{hpGR}
\ea
The reduction to the gravitational waveforms in GR is attributed 
to the absence of scalar radiation besides the exponential 
suppression of fifth forces outside compact objects 
in the mass range $m_s \gtrsim 10^{-11}$~eV.

\subsection{ppE parameters}

In the light mass regime $m_s \ll \omega$, the gravitational waveforms 
deviate from those in GR. 
Since the last terms in the square brackets of Eqs.~(\ref{Psiba})-(\ref{htf}) 
are the dominant terms arising from the modification of gravity, 
we ignore other correction terms such as $\delta_0$ and 
$\kappa_4 \Gamma^2$.
Then, one can express Eqs.~(\ref{hpf}) and (\ref{htf}) in the forms
\be
\tilde{h}_{\lambda}(f) \simeq \tilde{h}_{\lambda}^{\rm GR}(f)
\left[ 1-\frac{5 \kappa_4 (\hat{\alpha}_A-\hat{\alpha}_B)^2}
{48(G_* m \pi f)^{2/3}}
\right]e^{i\hat{\Psi}_{\lambda}}\,,
\label{thl}
\ee
where $\tilde{h}_{\lambda}^{\rm GR}(f)$ (with $\lambda=+,\times$) are
given in Eqs.~(\ref{htGR})-(\ref{hpGR}), and 
\ba
\hat{\Psi}_{\lambda} &\equiv&
\Psi_{\lambda}-\Psi_{\lambda}^{\rm GR} 
\simeq -\frac{5\kappa_4(\hat{\alpha}_A-\hat{\alpha}_B)^2}
{1792(G_* M_c \pi)^{5/3}(G_* m \pi)^{2/3} f^{7/3}}\,.
\label{hPsi}
\ea
These waveforms can be encompassed in the ppE 
framework \cite{Yunes:2009ke,Cornish:2011ys,Chatziioannou:2012rf,Tahura:2018zuq} 
given by 
\be
\tilde{h}_{\lambda}(f)=\tilde{h}_{\lambda}^{\rm GR}(f) 
\left[ 1+\sum_{j=1} \alpha_j 
\left( G_* M_c \pi f \right)^{a_j/3} \right] 
\exp \left[ i \sum_{j=1}\beta_j 
\left( G_* M_c \pi f \right)^{b_j/3} \right]\,,
\label{PPE}
\ee
where $\alpha_j$, $a_j$, $\beta_j$, and $b_j$
are constants parametrizing the deviation from GR. 
Comparing Eqs.~(\ref{thl})-(\ref{hPsi}) with Eq.~(\ref{PPE}), 
the ppE parameters in the light mass limit 
$m_s \ll \omega$ are given by 
\be
\alpha_1=-\frac{5}{48}\kappa_4 \left( \hat{\alpha}_A-\hat{\alpha}_B 
\right)^2 \eta^{2/5}\,,\qquad
a_1=-2\,,\qquad \beta_1=-\frac{5}{1792} \kappa_4
\left( \hat{\alpha}_A-\hat{\alpha}_B \right)^2 \eta^{2/5}\,,\qquad
b_1=-7\,,
\label{ppE2}
\ee
where 
\be
\eta \equiv \frac{\mu}{m}
=\left( \frac{M_c}{m} \right)^{5/3}\,.
\ee
In massless BD theories, the above ppE parameters 
reproduce those derived in 
Refs.~\cite{Chatziioannou:2012rf,Liu:2020moh}. 
For $\hat{\alpha}_A \neq \hat{\alpha}_B$, there are 
frequency-dependent corrections to the waveforms 
arising from the modification of gravity.

For the mass $m_s$ which is not much smaller than $\omega$, 
there are corrections to $\tilde{h}_{\lambda}(f)$ arising from the terms 
$m_s^2/\omega^2$. 
In this case, the second term in the square bracket of Eq.~(\ref{dome2}) 
is multiplied by the factor $(1-m_s^2/\omega^2)^{3/2} 
\simeq 1-3m_s^2/(2\omega^2)$. 
Then, the GW solution (\ref{thl}) with the phase (\ref{hPsi}) is modified to 
\ba
\tilde{h}_{\lambda}(f) &=& \tilde{h}_{\lambda}^{\rm GR}(f) 
\left[ 1-\frac{5\kappa_4 (\hat{\alpha}_A-\hat{\alpha}_B)^2}{48(G_* m \pi f)^{2/3}} 
\left( 1-\frac{3m_s^2}{2\pi^2 f^2} \right)\right]
e^{i\hat{\Psi}_{\lambda}}\,,\label{happ2} \\
\hat{\Psi}_{\lambda} &=& 
-\frac{5\kappa_4(\hat{\alpha}_A-\hat{\alpha}_B)^2}
{1792(G_* M_c \pi)^{5/3}(G_* m \pi)^{2/3} f^{7/3}}
\left( 1-\frac{105m_s^2}{208\pi^2 f^2} 
\right) \,. \label{hapt2} 
\ea
The leading-order ppE parameters are the same as those given in 
Eq.~(\ref{ppE2}). 
The light scalar mass $m_s$ gives rise to the following correction terms 
\be
\alpha_2=\frac{5}{32}\kappa_4 \left( \hat{\alpha}_A-\hat{\alpha}_B 
\right)^2 (G_* M_c  m_s)^2 \eta^{2/5}\,,\qquad
a_2=-8\,,\qquad \beta_2=\frac{15}{1664} \alpha_2\,,\qquad
b_2=-13\,.
\label{ppE4}
\ee
In the limit that $m_s \ll \pi f$, these corrections are negligibly 
small compared to the leading-order ppE contributions
given in Eq.~(\ref{ppE2}). 
In massive BD theories, the ppE parameters 
(\ref{ppE4}) coincide with those derived in Ref.~\cite{Liu:2020moh}.

\section{Application to concrete theories}
\label{contheory}

As we showed in the previous section, the ppE parameters 
crucially depend on $\hat{\alpha}_I$. In this section, we compute 
$\hat{\alpha}_I$ in concrete theories where the NS can have scalar hairs. 
In doing so, we first discuss an explicit relation between $\hat{\alpha}_I$ 
and a scalar charge by transforming the theory to an Einstein frame. 
The calculations of $\hat{\alpha}_I$ are important to probe the modification 
of gravity in strong gravity regimes in future observations
of GWs emitted from compact binaries.

\subsection{Nonminimally coupled theories 
and Einstein frame}
\label{nonmisec}

Let us consider theories given by 
the action (\ref{action}) with the nonminimal coupling 
$G_4(\phi)=\Mpl^2 F(\phi)/2$, i.e., 
\be
{\cal S}=
\int {\rm d}^4 x \sqrt{-g} 
\left[ \frac{\Mpl^2}{2}F(\phi)R+G_2(\phi,X)-G_3(\phi,X) \square \phi
\right]+{\cal S}_m (g_{\mu \nu}, \Psi_m)\,,
\label{action2}
\ee
which is known as the action in the Jordan frame where 
the matter fields $\Psi_m$ are minimally coupled to gravity.
To compute the quantities like $\hat{\alpha}_I$, it is convenient to 
perform a conformal transformation of 
the metric tensor as \cite{Fujii:2003pa,DeFelice:2010aj,Wald:1984rg}
\be
\hat{g}_{\mu \nu}=\Omega^2(\phi) g_{\mu \nu}\,,
\label{conformal}
\ee
where $\Omega^2 (\phi)$ is a field-dependent conformal factor, 
and a hat represents quantities in the transformed frame. 
To obtain the action in the Einstein frame, we use the following 
transformation properties
\be
\sqrt{-g}=\Omega^{-4} \sqrt{-\hat{g}}\,,\qquad 
R=\Omega^2 \left( \hat{R}+6\hat{\square}\omega-6 \hat{g}^{\mu \nu} 
\nabla_{\mu} \omega \nabla_{\nu} \omega \right)\,,\qquad
X=\Omega^2 \hat{X}\,,\qquad 
\square \phi=\Omega^2 \left( \hat{\square}\phi 
-2 \hat{g}^{\mu \nu} \nabla_{\mu}\omega \nabla_{\nu}\phi 
\right)\,,
\ee
where $\omega=\ln \Omega$. 
To transform the action (\ref{action2}) to that in the Einstein frame, 
we choose the conformal factor to be $\Omega^2(\phi)=F(\phi)$. 
Dropping boundary terms, the action in the Einstein frame is given by 
\be
\hat{S}=\int {\rm d}^4 x \sqrt{-\hat{g}} 
\left[ \frac{\Mpl^2}{2} \hat{R}+\hat{G}_2(\phi,\hat{X})
-\hat{G}_3(\phi,\hat{X}) \hat{\square} \phi
\right]+{\cal S}_m \left( F^{-1}(\phi)\hat{g}_{\mu \nu}, 
\Psi_m \right)\,,
\label{Ein}
\ee
where $\hat{X}=F^{-1}X$, and 
\be
\hat{G}_2=\frac{1}{F^2} \left[ G_2+F_{,\phi} \hat{X} 
\left( \frac{3}{2} \Mpl^2 F_{,\phi}-2G_3 \right) 
\right]\,,\qquad 
\hat{G}_3=\frac{G_3}{F}\,.
\label{tG2G3}
\ee
After the transformation, the matter fields $\Psi_m$ are 
coupled to the scalar field $\phi$ through the metric 
tensor $\hat{g}_{\mu \nu}$.

We will consider theories in which a standard kinetic term 
$\hat{X}$ is present in the Einstein frame. 
This is realized for the coupling 
function \cite{Kase:2020qvz,Minamitsuji:2022qku}
\be
G_2=\left( 1-\frac{3\Mpl^2 F_{,\phi}^2}{2F^2} 
\right)F(\phi) X+K(\phi,X)\,,
\ee
where $K$ is a function of $\phi$ and $X$. 
Then, it follows that 
\be
\hat{G}_2=\hat{X}+\frac{K}{F^2}-\frac{2F_{,\phi}}{F^2}
G_3 \hat{X}\,,\qquad 
\hat{G}_3=\frac{G_3}{F}\,.
\ee

We can further specify theories containing a quadratic 
kinetic term $\mu_2 X^2$ and a scalar potential $V(\phi)$ 
in $K$, such that $K=\mu_2 X^2-V(\phi)$.
Taking the cubic Galileon term $G_3=\mu_3 X$ into 
account as well, the coupling functions in the Jordan 
frame yield
\be
G_2=\left( 1-\frac{3\Mpl^2 F_{,\phi}^2}{2F^2} 
\right)F(\phi) X+\mu_2 X^2-V(\phi)\,,\qquad G_3=\mu_3 X\,,
\qquad G_4(\phi)=\frac{\Mpl^2}{2}F(\phi)\,,
\label{G234}
\ee
where $\mu_2$ and $\mu_3$ are constants.
In the Einstein frame, the coupling functions $\hat{G}_2$ 
and $\hat{G}_3$ yield 
\be
\hat{G}_2=\hat{X}+\mu_2 \hat{X}^2
-\frac{2F_{,\phi}}{F}\mu_3 \hat{X}^2-\frac{V}{F^2}\,,\qquad 
\hat{G}_3=\mu_3 \hat{X}\,.
\ee

In the following, we present theories that belong to 
the action (\ref{action2}) with the coupling functions (\ref{G234}).
\begin{itemize}
\item (i) BD theories with a scalar potential 
$V(\phi)$ \cite{Brans:1961sx}:
\be
G_2=(1-6Q^2)e^{-2Q\phi/\Mpl} X-V(\phi)\,,\qquad 
G_3=0\,,\qquad
G_4=\frac{\Mpl^2}{2}e^{-2Q\phi/\Mpl}\,,
\label{BDaction}
\ee
where the nonminimal coupling corresponds to 
$F(\phi)=e^{-2Q\phi/\Mpl}$, and $Q$ is a constant 
characterizing the coupling strength between 
the scalar field and gravity sector.
Setting $\chi=F=e^{-2Q\phi/\Mpl}$ with $\Mpl=1$, 
it follows that the above theory is equivalent to the 
action ${\cal S}=\int {\rm d}^4 x \sqrt{-g}\, 
[\chi R/2-\omega_{\rm BD}\nabla^{\mu}\chi \nabla_{\mu}\chi/
(2\chi)-V]+{\cal S}_m$ originally propsed by 
Brans and Dicke \cite{Brans:1961sx}.
Here, the BD parameter $\omega_{\rm BD}$ is related to 
the coupling $Q$ according to \cite{Khoury:2003rn,Tsujikawa:2008uc}
\be
3+2\omega_{\rm BD}=\frac{1}{2Q^2}\,.
\label{BDpara}
\ee
GR corresponds to the limit $\omega_{\rm BD} 
\to \infty$, i.e., $Q \to 0$. 
Metric $f(R)$ gravity with the action 
${\cal S}=\int {\rm d}^4 x \sqrt{-g}\, \Mpl^2f(R)/2$
belongs to a subclass of BD theories given by the coupling 
functions (\ref{BDaction}), with the correspondence 
$Q=-1/\sqrt{6}$, $V(\phi)=\Mpl^2 ( FR-f )/2$, and  
$F=\partial f/\partial R=e^{-2Q\phi/\Mpl}$ \cite{Amendola:2006kh,DeFelice:2010aj}. 
If the mass of $\phi$ is as light as today's Hubble 
constant $H_0$ at low redshifts, the scalar field $\phi$ 
can be also the source for 
dark energy \cite{Amendola:1999qq,Bartolo:1999sq,Boisseau:2000pr,Tsujikawa:2008uc,Tsujikawa:2019pih}.

\item (ii) Theories with spontaneous 
scalarization \cite{Damour:1993hw,Damour:1996ke}:
\be
G_2=\left( 1-\frac{3\Mpl^2 F_{,\phi}^2}{2F^2} 
\right)F(\phi) X\,,\qquad 
G_3=0\,,\qquad 
G_4=\frac{\Mpl^2}{2}F(\phi)\,,
\ee
where $F$ is a function containing the even power-law dependence 
of $\phi$. The nonminimal coupling chosen by 
Damour and Esposite-Farase is given by 
$F(\phi)=e^{-\beta \phi^2/(2\Mpl^2)}$, 
where $\beta$ is a constant. 
On the static and spherically symmetric background,  there is 
a nonvanishing scalar-field 
branch $\phi (r) \neq 0$ besides the GR branch $\phi(r) = 0$, 
where $r$ is the distance from the center of symmetry. 
For strong gravitational stars like the NS, the necessary condition 
for the occurrence of spontaneous scalarization from the GR branch 
to the other branch is given by $F_{,\phi \phi}(0)>0$, 
which translates to the condition $\beta<0$. 
If we apply this model to cosmology, there is tachyonic growth of 
$\phi$ that can violate the dynamics of successful cosmic expansion 
history \cite{Damour:1992kf,Damour:1993id}. 
This problem is circumvented by the presence of a coupling 
$g^2 \phi^2 \chi^2/2$ between $\phi$ and 
an inflaton field $\chi$ \cite{Anson:2019ebp,Nakarachinda:2022tjj}, 
in which case $\phi$ is exponentially suppressed during inflation.
Note that this coupling does not affect the mechanism of 
spontaneous scalarization because of the decay of $\chi$ 
by the end of reheating.

\item (iii) Scalarized NSs with a scalar potential $V(\phi)$ 
and a positive nonminimal coupling constant $\beta>0$ \cite{Minamitsuji:2022qku}:
\be
G_2=\left( 1-\frac{3\beta^2\phi^2}{2\Mpl^2} 
\right)e^{-\beta \phi^2/(2\Mpl^2)} X-V(\phi)\,,\qquad 
G_3=0\,,\qquad 
G_4=\frac{\Mpl^2}{2}e^{-\beta \phi^2/(2\Mpl^2)}\,.
\label{symbre}
\ee
The difference from original spontaneous scalarization is that 
there is a self-interacting potential of the type
\be
V(\phi)=m_s^2 f_B^2 \left[ 1+\cos \left( \frac{\phi}{f_B} 
\right) \right]\,,
\label{Vphi}
\ee
where $m_s$ and $f_B$ are constants. 
Far away from the NS, the scalar field sits at the
vacuum expectation value $\phi_v=\pi f_B$. 
Inside the NS, a nonminimal coupling with $\beta>0$ can 
dominate over a negative mass squared of the bare potential $-m_s^2$.
This leads to the symmetry restoration with the central field value 
$\phi_c$ close to $0$. Then, there are scalarized NS solutions 
connecting $\phi_v$ with $\phi_c$ whose difference 
is significant on strong gravitational backgrounds 
(see Ref.~\cite{Babichev:2022djd} for a model of scalarized BHs 
based on a scalar-Gauss-Bonnet coupling). 
In this scenario, the scalar field is not subject to tachyonic instability during 
inflation and it finally approaches a vacuum expectation value 
without spoiling a successful cosmological 
evolution \cite{Minamitsuji:2022qku}. 
\end{itemize}

In Refs.~\cite{Chagoya:2014fza,Ogawa:2019gjc}, the authors took into account 
the cubic Galileon coupling $G_3=\mu_3 X$ for the theories of 
types (i) and (ii) and showed that the deviation from GR is suppressed 
even for relativistic stars through the Vainshtein mechanism. 
If the Vainshtein radius $r_V$ is much larger than the radius $r_s$  
of star, nonlinear scalar derivative terms like 
$(\square_{\rm M} \varphi)^2$ and 
$\partial^{\mu}\partial^{\nu} \varphi
\partial_{\mu}\partial_{\nu} \varphi$ dominate over 
$\square_{\rm M} \varphi$ at the distance $r<r_V$. 
For $r_V \gg r_s$, the computation of the gravitational waveform based on 
the PN expansion (\ref{phieS}) outside the star loses 
its validity inside the Vainshtein radius.
If $r_V$ is at most of order $r_s$, i.e., $r_V \lesssim r_s$, the PN solutions 
outside the star are still valid. 
In this latter situation, the screening of fifth forces should occur only 
inside the star. In this case, the cubic coupling constant $\mu_3$ 
needs to be tuned to realize $r_V$ same order as $r_s={\cal O}(10~{\rm km})$. 
We will not address such a specific case in this paper. 

Instead, we study the effect of the $\mu_2 X^2$ term on 
$\hat{\alpha}_I$ by setting $\mu_3=0$ in Eq.~(\ref{G234}). 
We also consider the case in which 
the scalar potential is absent, so that the coupling functions 
in the Jordan frame are 
\be
G_2=\left( 1-\frac{3\Mpl^2 F_{,\phi}^2}{2F^2} 
\right)F(\phi) X+\mu_2 X^2\,,\qquad G_3=0\,,
\qquad G_4(\phi)=\frac{\Mpl^2}{2}F(\phi)\,.
\label{G234d}
\ee
In the Einstein frame, the action is given by Eq.~(\ref{Ein}) with 
\be
\hat{G}_2=\hat{X}+\mu_2 \hat{X}^2\,,\qquad 
\hat{G}_3=0\,.
\ee
In this class of theories, there are no asymptotically flat 
hairy BH solutions known in the 
literature \cite{Hawking:1972qk,Bekenstein:1995un,Sotiriou:2011dz,Graham:2014mda,Faraoni:2017ock,Minamitsuji:2022vbi}. 
Thus, we only consider a static and spherically symmetric NS to 
compute the quantities appearing in Eq.~(\ref{ppE2}).

\subsection{How to compute $\hat{\alpha}_I$}

The line element corresponding to the static 
and spherically symmetric background in the 
Jordan frame is given by 
\be
{\rm d}s^2
=-f(r) \rd t^{2} +h^{-1}(r) \rd r^{2}
+ r^{2} \rd \Omega^2\,,
\label{metric}
\ee
where $f(r)$ and $h(r)$ are functions of the radial
coordinate $r$. 
For the matter fields $\Psi_m$ inside the NS, we consider a perfect fluid given by 
the energy-momentum tensor ${T^{\mu}}_{\nu}={\rm diag}[-\rho(r), P(r), P(r), P(r)]$ 
minimally coupled to gravity, 
where $\rho$ is the density and $P$ is the pressure.
On the background given by the line element (\ref{metric}), 
the field equations of motion 
are \cite{Kobayashi:2012kh,Kobayashi:2014wsa,Kase:2020qvz,Minamitsuji:2016hkk,Kase:2021mix}
\ba
& &
\frac{f'}{f}=-\frac{F^2[4\Mpl^2 (h-1)-2h r^2 \phi'^2]
+3\Mpl^2 r^2 h \phi'^2 F_{,\phi}^2+rF[h \phi' 
(8 F_{,\phi} \Mpl^2+3\mu_2 rh \phi'^3)-4rP]}
{2F (2F+F_{,\phi}r \phi')\Mpl^2 rh}\,,\label{eqmo1}\\
& &
\frac{h'}{h}-\frac{f'}{f}=-r \frac{2F^2 h \phi'^2+2F[h \Mpl^2 (F_{,\phi \phi} \phi'^2
+F_{,\phi}\phi'')-\mu_2 h^2 \phi'^4+\rho+P]-3\Mpl^2 F_{,\phi}^2 h \phi'^2}
{F (2F+F_{,\phi}r \phi')\Mpl^2 h}\,,\\
& &
\frac{1}{r^2} \sqrt{\frac{h}{f}} \left( r^2 \sqrt{\frac{f}{h}} J^r \right)'
+{\cal P}_{\phi}=0\,,\\
& &
P'+\frac{f'}{2f} (\rho+P)=0\,,\label{eqmo4}
\ea
where a prime represents the derivative with respect to $r$, and 
\ba
\hspace{-0.9cm}
J^r &=& h \phi' \left( F-\frac{3\Mpl^2 F_{,\phi}^2}{2F} 
-\mu_2 h \phi'^2 \right)\,,\\
\hspace{-0.9cm}
{\cal P}_{\phi} &=& \frac{F_{,\phi}}{4} \left[ \frac{\Mpl^2 \{ 
r^2 h f'^2-4f^2 (rh'+h-1)-rf(2rh f''+rf'h'+4hf') \}}{r^2 f^2}
-\frac{h \phi'^2 \{ 2F^2+3\Mpl^2 (F_{,\phi}^2-2F F_{,\phi \phi}) \}}{F^2} 
\right].
\ea
The Arnowitt-Deser-Misner (ADM) mass $m$ of the star is related 
to the metric component $h$ as 
\be
m=4 \pi \Mpl^2 r \left[ 1-h(r) \right]|_{r \to \infty}\,.
\ee
At $r=0$, we impose the regular boundary conditions 
$f(0)=f_c$, $h(0)=1$, $\phi(0)=\phi_c$, $\rho(0)=\rho_c$, 
$P(0)=P_c$, and $f'(0)=h'(0)=\phi'(0)=\rho'(0)=P'(0)=0$. 
Then, the solutions expanded around the center of star consistent with 
these boundary conditions are 
\ba
f &=& f_c+\frac{f_c}{6F(\phi_c)\Mpl^2} 
\left[ \rho_c+3P_c +\frac{\Mpl^2 F_{,\phi}^2 (\phi_c) 
(\rho_c-3P_c)}{2F^2(\phi_c)} \right]r^2
+{\cal O}(r^4)\,,\label{fr=0} \\
h &=& 1-\frac{1}{3F(\phi_c)\Mpl^2} \left[ \rho_c-
\frac{\Mpl^2 F_{,\phi}^2 (\phi_c) 
(\rho_c-3P_c)}{2F^2(\phi_c)} \right]r^2
+{\cal O}(r^4)\,,\\
\phi &=& \phi_c-\frac{F_{,\phi}(\phi_c) (\rho_c-3P_c)}
{12F^2(\phi_c)}r^2+{\cal O}(r^4)\,,\label{phicen} \\
P &=& P_c-\frac{(\rho_c+P_c)[2F^2(\phi_c)(\rho_c+3P_c)
+F_{,\phi}^2(\phi_c)\Mpl^2 (\rho_c-3P_c)]}{24F^3 (\phi_c) \Mpl^2}r^2
+{\cal O}(r^4)\,.\label{Pr=0}
\ea
{}From Eq.~(\ref{phicen}) it is clear that the hairy NS solution arises 
through the coupling between the scalar field and matter 
mediated by the nonminimal coupling $F(\phi)R$. 
For larger values of $|F_{,\phi}(\phi_c)|$ and $|\rho_c-3P_c|$, 
the derivative $|\phi'(r)|$ tends to be larger. 
The contribution of the term $\mu_2 X^2$ appears 
at the order of $r^4$ in Eqs.~(\ref{fr=0})-(\ref{Pr=0}). 
Since $|\phi'(r)|$ grows as a function of $r$ inside the star, 
the higher-order term $\mu_2 X^2$ can also contribute
to the solutions around $r=r_s$.

We define the radius of star $r_s$ according to the condition $P(r_s)=0$ 
and assume that $\rho=0=P$ in the exterior region of star. 
To study the scalar-field solution for $r>r_s$, 
it is convenient to transform the metric to that 
in the Einstein frame such that
\be
{\rm d}\hat{s}^2
=F(\phi){\rm d}s^2
=-\hat{f}(\hat{r}) \rd t^{2} +\hat{h}^{-1}(\hat{r}) \rd \hat{r}^{2}
+ \hat{r}^{2} \rd \Omega^2\,,
\label{metEin}
\ee
where $\hat{r}$, $\hat{f}$, and $\hat{h}$ are related to those 
in the Jordan frame as  
\be
\hat{r}=\sqrt{F}\,r\,,\qquad 
\hat{f}=F f\,,\qquad 
\hat{h}= \left( 1+\frac{F_{,\phi}\phi'}{2F}r \right)^2h\,.
\label{rmco}
\ee
The fluid density $\hat{\rho}$, pressure $\hat{P}$, and 
ADM mass $\hat{m}_I$ of the NS (labeled by $I$) 
in the Einstein frame 
are expressed as
\be
\hat{\rho}=\frac{\rho}{F^2}\,,\qquad 
\hat{P}=\frac{P}{F^2}\,,\qquad
\hat{m}_I=\frac{m_I}{\sqrt{F}}\,.
\label{rmco2}
\ee
In the Einstein frame, the scalar-field equation of motion is 
written in the form 
\be
\frac{1}{\hat{r}^2} \sqrt{\frac{\hat{h}}{\hat{f}}} 
\frac{\rd}{\rd \hat{r}} \left\{ \left[ 1-\mu_2 \hat{h} 
\left( \frac{\rd \phi}{\rd \hat{r}} \right)^2 \right] \hat{r}^2 \sqrt{\hat{f} \hat{h}}
\frac{\rd \phi}{\rd \hat{r}} \right\}=-\frac{F_{,\phi}}{2F} 
\left( \hat{\rho}-3\hat{P} \right)\,.
\label{scaEin}
\ee
Since $\hat{\rho}=0=\hat{P}$ outside the NS, 
Eq.~(\ref{scaEin}) is integrated to give 
\be
\left[ 1-\mu_2 \hat{h} \left( \frac{{\rm d}\phi}{{\rm d}\hat{r}} 
\right)^2 \right] 
\frac{{\rm d}\phi}{{\rm d}\hat{r}}
=\frac{q_s}{\hat{r}^2 \sqrt{\hat{f}\hat{h}}}\,,
\ee
where $q_s$ is a constant corresponding to a scalar charge. 
At spatial infinity ($\hat{r} \to \infty$), we impose the asymptotically flat 
boundary conditions ${\rm d} \phi/{\rm d}\hat{r} \to 0$, 
$\hat{f} \to 1$, and $\hat{h} \to 1$.  
Then, far away from the star, the scalar field has the
following asymptotic behavior
\be
\frac{{\rm d}\phi}{{\rm d}\hat{r}}=\frac{q_s}{\hat{r}^2}\,,\qquad 
\phi (\hat{r})=\phi_0-\frac{q_s}{\hat{r}}\,,
\label{phila}
\ee
where $\phi_0$ is the asymptotic value of $\phi$. 
The higher-order kinetic term $\mu_2 \hat{X}^2$ is 
suppressed in this regime, so that the canonical kinetic term $\hat{X}$ gives 
a dominant contribution to the ADM mass $\hat{m}_I$ in the Einstein frame. 
In other words, $\hat{m}_I$ acquires the $\phi$ dependence through the 
Lagrangian $L_{\phi}=\hat{X}=-(1/2)\hat{g}^{ij} 
\partial_{i} \phi \partial_{j} \phi$, where $L_{\phi}$ does not 
contain the time dependence of $\phi$ on the static 
background (\ref{metEin}).
Since $L_{\phi}$ contributes to $\hat{m}_I (\phi)$ through 
the three dimensional volume integral $-\int {\rm d}^3 x\,L_{\phi}$, 
varying $\hat{m}_I (\phi)$ with respect to $\phi$ 
leads to \cite{Damour:1992we}
\be
\delta \hat{m}_I (\phi)=-\int {\rm d}^3 x\,\delta L_{\phi}
=-\int {\rm d}^3 x\,\partial_{i} \left[ \frac{\partial L}
{\partial (\partial_{i}\phi)} \delta \phi \right]
=-\int {\rm d}^2 S_{i} \frac{\partial L}
{\partial (\partial_{i}\phi)}\delta \phi
=\int {\rm d}^2 S_i\,\partial^{i}\phi\,\delta \phi\,,
\ee
where in the second equality we used the 
Euler-Lagrange equation, and in the third equality 
the volume integral is changed to the surface integral 
upon using the Gauss's theorem. 
Then, it follows that 
\be
\hat{m}_{I,\phi}=\int {\rm d}^2 S_i \partial^{i}\phi
=4\pi \hat{r}^2 \frac{q_s}{\hat{r}^2}=4\pi q_s\,.
\ee
On using the correspondence $\hat{m}_I=m_I/\sqrt{F}$, 
the quantity $\hat{\alpha}_I$ defined in Eq.~(\ref{halI}) 
can be expressed as\footnote{Damour and Esposite-Farese \cite{Damour:1993hw} introduced a dimensionless scalar field 
$\phi_{\rm DEF}=\phi/(\sqrt{2}\Mpl)$ and defined the quantity
$\hat{\alpha}_I^{\rm DEF}={\rm d} \ln \hat{m}_I/{\rm d} \phi_{\rm DEF}$. 
Hence $\hat{\alpha}_I^{\rm DEF}$ is $\sqrt{2}$ times as large as 
our definition of $\hat{\alpha}_I$, i.e., 
$\hat{\alpha}_I^{\rm DEF}=\sqrt{2}\hat{\alpha}_I$.}
\be
\hat{\alpha}_I = \Mpl \frac{{\rm d} \ln \hat{m}_I(\phi)}
{{\rm d} \phi}\biggr|_{\phi=\phi_0}\,.
\ee
Then, we obtain the following relation
\be
q_s=\frac{\hat{m}_I}{4\pi \Mpl} \hat{\alpha}_I\,,
\ee
which shows that $\hat{\alpha}_I$ is directly related to 
the scalar charge $q_s$. 
It is worth mentioning that the quantity $\alpha_I$ defined 
in the Jordan frame does not correspond the scalar charge 
due to the presence of the nonminimal coupling $G_4(\phi)R$.  
At large distances, the scalar-field solution (\ref{phila}) 
is expressed as 
\be
\phi(\hat{r})=\phi_0-\frac{\hat{m}_I \hat{\alpha}_I}{4\pi \Mpl \hat{r}}\,.
\label{phiE}
\ee
On using Eqs.~(\ref{rmco})-(\ref{rmco2}), 
we can write Eq.~(\ref{phiE}) in terms of the quantities in the Jordan frame as
\be
\phi(r)=\phi_0-\frac{m_I \hat{\alpha}_I}{4\pi F(\phi_0) \Mpl r}\,.
\label{phila2}
\ee

To estimate the values of $\hat{\alpha}_I$, we extrapolate the two asymptotic 
solutions (\ref{phicen}) and (\ref{phila2}) up to the distance $r=r_s$ and 
match the $r$ derivatives of them at $r=r_s$, i.e., 
\be
-\frac{F_{,\phi}(\phi_c) \rho_c (1-3w_c)}
{6F^2(\phi_c)}r_s \simeq \frac{m_I \hat{\alpha}_I}
{4\pi F(\phi_0) \Mpl r_s^2}\,,
\ee
where $w_c=P_c/\rho_c$ is the fluid equation of state (EOS) parameter 
at $r=0$. For a star with a nearly constant density $\rho_c$, we can 
use the approximation $m_I \simeq 4\pi r_s^3\rho_c/3$. 
Exploiting the approximation $F(\phi_c) \simeq F(\phi_0)$ further, 
the order of $\hat{\alpha}_I$ can be estimated as 
\be
\hat{\alpha}_I \simeq -\frac{g_4(\phi_c)}{2}
(1-3w_c)\,,
\label{halpha}
\ee
where 
\be
g_4(\phi_c)=\frac{\Mpl F_{,\phi}(\phi_c)}{F(\phi_c)}\,.
\ee
This shows that $\hat{\alpha}_I$ is related to the $\phi$ 
dependence of nonminimal couplings at $r=0$. 
In the limit of point-like sources, i.e., $r_s \to 0$, 
the relation (\ref{halpha}) becomes exact.
For nonrelativistic stars ($w_c \ll 1$), we have 
$\hat{\alpha}_I \simeq -g_4(\phi_c)/2$. 
For NSs, $w_c$ can be of order 0.1 and hence the approach to the value 
$w_c=1/3$ tends to suppress $\hat{\alpha}_I$.

The approximate formula (\ref{halpha}) is valid for both relativistic 
and nonrelativistic stars, but it cannot be applied to BHs. Indeed, 
it is known that there are no asymptotically flat hairy BH 
solutions with $\hat{\alpha}_I \neq 0$ in theories under consideration.
Then, for the BH-BH binary system, we have $\hat{\alpha}_A=0=\hat{\alpha}_B$ 
and hence the ppE parameters are not modified in comparison to GR. 
On the other hand, the NS can have scalar hairs for theories like 
(i)-(ii) mentioned in Sec.~\ref{nonmisec}. 
As we will see in Sec.~\ref{theosec} in concrete theories, 
the values of $\hat{\alpha}_I$
are different depending on the ADM mass of NSs. 
Provided that there are some mass difference between two NSs, 
it is possible to place constraints on the difference 
$\hat{\alpha}_A-\hat{\alpha}_B$ from the gravitational 
waveform emitted from the NS-NS binary.

For the NS-BH binary system, the difference between 
the scalarized NS ($\hat{\alpha}_A \neq 0$) and the no-hair 
BH ($\hat{\alpha}_B=0$) can be generally larger than that of the NS-NS 
binary. We will focus on such a case in the following discussion.
The ppE parameter $\beta_1$ can be constrained from the 
phase of observed gravitational waveforms emitted from the NS-BH binary. 
If the GW observations give the bound $|\beta_1| \le |\beta_1|^{\rm max}$, 
it translates to
\be
|\hat{\alpha}_A| \le 16 \sqrt{\frac{7}{5}|\beta_1|^{\rm max}}
\left( \frac{m}{\mu} \right)^{1/5}
\sqrt{\frac{\zeta_0 \Mpl^2}{G_4(\phi_0)}}\,.
\ee
In this way, we can constrain $\hat{\alpha}_A$ 
from the GW observations.

Before computing $\hat{\alpha}_A$ in concrete theories, we discuss 
the stability of NSs against odd- and even-parity perturbations 
on the static and spherically symmetric background.
For theories given by the coupling functions (\ref{G234d}) 
in the Jordan frame, there are neither ghost nor 
Laplacian instabilities under the following 
three conditions \cite{Kase:2021mix}
\be
G_4>0\,,\qquad G_{2,X}G_4+3G_{4,\phi}^2>0\,,\qquad 
G_{2,X}G_4+3G_{4,\phi}^2+2X G_{2,XX}G_4>0\,,
\ee
which translate, respectively, to 
\be
F>0\,,\qquad 
\mu_2 h \phi'^2<F\,,\qquad 
3\mu_2 h \phi'^2<F\,.
\ee
The first inequality is ensured for the choices of nonminimal 
couplings like $F=e^{-2Q \phi/\Mpl}$ and 
$F=e^{-\beta \phi^2/(2\Mpl^2)}$.
For $\mu_2<0$, the second and third conditions are 
automatically satisfied. For $\mu_2>0$, 
the second and third inequalities give the condition
\be
3\mu_2 h \phi'^2<F\,,
\label{nogho}
\ee
so that the positive coupling $\mu_2$ is bounded from above.
With the condition $F>0$, Eq.~(\ref{nogho}) translates to 
\be
3\mu_2 \hat{h} \left( \frac{\rd \phi}{\rd \hat{r}} \right)^2<1\,,
\label{nogho2}
\ee
which corresponds to the stability condition 
in the Einstein frame.

\subsection{Concrete theories}
\label{theosec}

\subsubsection{BD theories with $G_2 \supset \mu_2 X^2$}

Let us first consider theories given by the 
nonminimal coupling 
\be
F(\phi)=e^{-2Q \phi/\Mpl}\,,
\label{nonmi1}
\ee
with the coupling functions (\ref{G234d}). 
For $\mu_2=0$, this is equivalent to massless BD theory, 
where the BD parameter $\omega_{\rm BD}$ is related to 
the coupling $Q$ as Eq.~(\ref{BDpara}). 
As in the case of solutions (\ref{phicen}) and (\ref{phila2}) 
derived for NSs, stars on the weak gravity background acquire the scalar 
charge through the nonminimal coupling as well. 
This mediates fifth forces in the solar system, 
which are constrained by local gravity experiments.
The solar-system tests of gravity have placed the bound 
$\omega_{\rm BD}>4.0 \times 10^4$ \cite{Will:2014kxa}, 
which translates to the upper limit 
\be
|Q|<2.5 \times 10^{-3}\,.
\label{Qcon}
\ee
For the nonminimal coupling (\ref{nonmi1}), the NS has a scalar charge 
given by Eq.~(\ref{halpha}), i.e., 
\be
\hat{\alpha}_A \simeq Q (1-3w_c)\qquad \quad 
({\rm for}~\mu_2=0).
\label{alphaes1}
\ee
If we consider nonrelativistic point-like sources ($w_c \to 0$ and $r_s \to 0$), 
Eq.~(\ref{alphaes1}) gives the exact relation $\hat{\alpha}_A=Q$. 
In this case, Eq.~(\ref{q}) yields
\be
\tilde{G}=\frac{G}{F} \left( 1+2Q^2 \right)
=\frac{G}{F} \frac{4+2\omega_{\rm BD}}{3+2\omega_{\rm BD}} \qquad \quad 
({\rm for}~\mu_2=0,~w_c \to 0,~r_s \to 0),
\label{Gweak}
\ee
where $G=(8 \pi \Mpl^2)^{-1}$, and we used the relation (\ref{BDpara}) 
in the second equality.
Thus, in massless BD theories, the effective gravitational coupling between two nonrelativistic point-like sources is given by Eq.~(\ref{Gweak}).

For NSs, $w_c$ can be of order 0.1. To compute $\hat{\alpha}_A$ for 
NSs with finite radius $r_s$, 
we also need to consider realistic EOSs inside the star ($0 \le r \le r_s$) 
without approximating it as a point particle. 
We numerically integrate Eqs.~(\ref{eqmo1})-(\ref{eqmo4}) from a central 
region of the star to a sufficiently large distance by specifying a NS EOS and 
compute $\hat{\alpha}_A$ by comparing numerical solutions of 
$\phi$ with the asymptotic solution (\ref{phila2}). 
For $\mu_2=0$ the similar analysis was already performed 
in the literature \cite{Niu:2021nic}, so we will not repeat it here. 
Numerically, we confirmed that the approximate formula 
(\ref{alphaes1}) gives a good criterion for the estimation of $\hat{\alpha}_A$ 
with the EOS parameter $w_c$ in the range $w_c<1/3$. 
{}From the solar-system constraint (\ref{Qcon}), the parameter 
$\hat{\alpha}_A$ should be in the range 
$|\hat{\alpha}_A|<0.0025 (1-3w_c)$. 
If we consider a NS-BH binary with the masses $m_A=1.7 M_{\odot}$ 
and $m_B=2.5 M_{\odot}$, for example, the scalar charge 
$|\hat{\alpha}_A|=0.002$ corresponds 
to the ppE parameter $|\beta_1|$ of order $10^{-9}$. 
If the future GW observations can pin down the value $|\beta_1|^{\rm max}$ 
to the order $10^{-9}$, it is possible to obtain 
tighter bounds on $|Q|$ than those constrained from the solar-system 
tests of gravity.

For $\mu_2 \neq 0$, the higher-order derivative term $\mu_2 X^2$ 
can contribute to solutions of $\phi$, $f$, and $h$ around the 
surface of star. We are interested in the case where the effect of 
$\mu_2 X^2$ becomes important in strong gravity regimes,  
while it is suppressed relative to $X$ in the solar system.
In other words, we search for the possibility for 
enhancing $|\hat{\alpha}_A|$ by the derivative 
term $\mu_2 X^2$ relative to (\ref{alphaes1}), 
while respecting the bound (\ref{Qcon}). 
For this purpose, we write the scalar-field Eq.~(\ref{scaEin}) explicitly as 
\be
\left( 1-3\mu_2 \hat{h} \hat{\phi}'^2 \right)\hat{h} 
\hat{\phi}''+\left( 1-\mu_2 \hat{h} \hat{\phi}'^2 \right)\hat{h} 
\left( \frac{2}{\hat{r}}+\frac{\hat{f}'}{2\hat{f}}
+\frac{\hat{h}'}{2\hat{h}} \right) \hat{\phi}'
-\mu_2  \hat{h} \hat{h}' \hat{\phi}'^3
=-\frac{F_{,\phi}}{2F} 
\left( \hat{\rho}-3\hat{P} \right)\,,
\label{fieledeqEin}
\ee
where we used the notations $\hat{\phi}'=\rd \phi/\rd \hat{r}$ and 
$\hat{\phi}''=\rd^2 \phi/\rd \hat{r}^2$. 
A prime here represents the derivative with respect to $\hat{r}$.
A positive coupling $\mu_2$ gives the coefficient 
$1-3\mu_2 \hat{h} \hat{\phi}'^2$ 
smaller than 1, so it may be possible to enhance the overall amplitude 
of $|\hat{\phi}'|$. Unless the term $1-3\mu_2 \hat{h} \hat{\phi}'^2$ 
is very close to 0 or is negative, however,  
we numerically find that $\hat{\alpha}_A$
is practically the same as that for $\mu_2=0$. 
Since the stability of NSs requires the condition 
(\ref{nogho2}), the coupling $\mu_2$ with $1-3\mu_2 \hat{h} \hat{\phi}'^2<0$ 
is excluded. When the term $1-3\mu_2 \hat{h} \hat{\phi}'^2$ is very close to 0, 
there is a strong coupling problem associated with a small coefficient 
of the second derivative $\hat{\phi}''$ in Eq.~(\ref{fieledeqEin}). 
Hence it is not possible to realize a value of 
$|\hat{\alpha}_A|$
whose order is larger than (\ref{alphaes1}). 
We note that the negative coupling $\mu_2$ tends to suppress 
$|\hat{\phi}'|$, so $|\hat{\alpha}_A|$ does not exceed the value 
for $\mu_2=0$.

\subsubsection{Spontaneous scalarization with $G_2 \supset \mu_2 X^2$}

For $\mu_2=0$, spontaneous scalarization of NSs can be realized by 
nonminimal couplings containing the even power-law dependence of $\phi$.
A typical example is given by the coupling 
function \cite{Damour:1993hw,Damour:1996ke}
\be
F(\phi)=e^{-\beta \phi^2/(2\Mpl^2)}\,,
\ee
which allows the existence of a nontrivial branch $\phi(r) \neq 0$ 
besides the GR branch $\phi(r)=0$. 
On the strong gravitational background, 
the GR branch can be subject to tachyonic instability for 
$\beta<0$, which is triggered by spontaneous growth of $\phi$ toward the other nontrivial branch.
Spontaneous scalarization of NSs is a nonperturbative phenomenon which can occur for largely negative couplings in the range 
$\beta \le -4.35$ \cite{Harada:1998ge,Novak:1998rk,Sotani:2004rq,Silva:2014fca,Barausse:2012da}.

{}From Eq.~(\ref{halpha}), the scalar charge can be estimated as 
\be
\hat{\alpha}_I \simeq \frac{\beta \phi_c}{2\Mpl}
(1-3w_c)\,.
\label{alI}
\ee
Since $\phi_c$ is nonvanishing for the scalarized branch, 
NSs acquire the scalar charge. 
The asymptotic field value $\phi_0$ at spatial infinity 
is constrained by solar-system tests of gravity. 
Since the parametrized PN parameter in the current theory is  
$\gamma_{\rm PPN}-1=
-2\beta^2 \phi^2/(2\Mpl^2+\beta^2 \phi^2)$ \cite{Damour:1992we}, 
the constraint $\gamma_{\rm PPN}-1=(2.1 \pm 2.3) \times 10^{-5}$ 
arising from the Shapiro time delay experiment \cite{Will:2014kxa}
gives the upper limit
\be
|\phi_0| \le 1.4 \times 10^{-3} \Mpl |\beta|^{-1}\,.
\label{phi0}
\ee
For given model parameters, we iteratively search for a central field value 
$\phi_c$ consistent with the bound (\ref{phi0}).
To describe realistic nuclear interactions inside NSs, we exploit 
an analytic representation of the SLy EOS given in Ref.~\cite{Haensel:2004nu}.
For the numerical purpose, we introduce the following 
constants
\be
\rho_0=1.6749 \times 10^{14}~{\rm g/cm}^3\,,\qquad 
r_0=\sqrt{\frac{8\pi \Mpl^2}{\rho_0}}=89.664~{\rm km}\,,
\ee
which are used to normalize $\rho$ and $r$, respectively.

\begin{figure}[ht]
\begin{center}
\includegraphics[height=3.2in,width=3.4in]{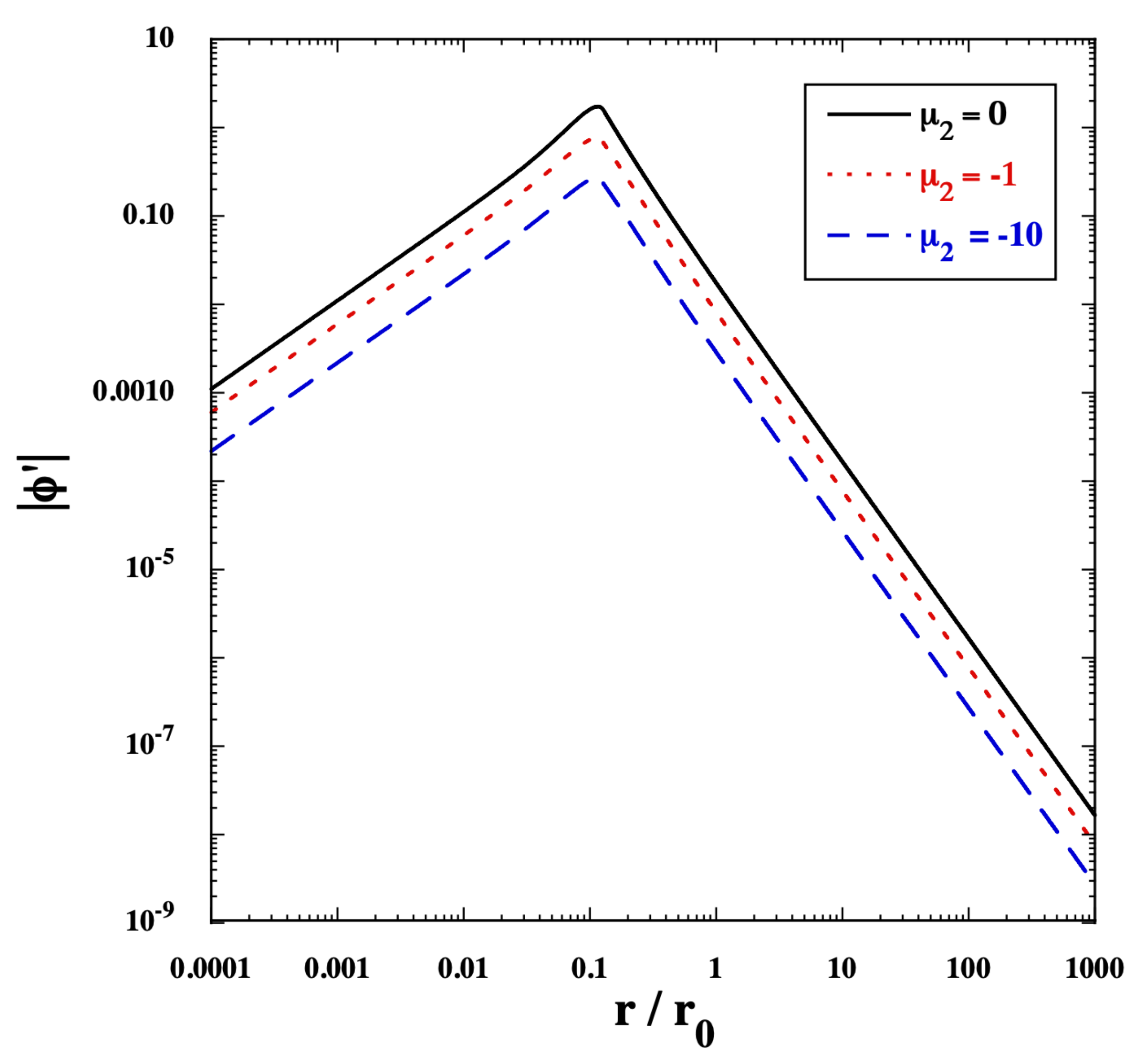}
\includegraphics[height=3.2in,width=3.5in]{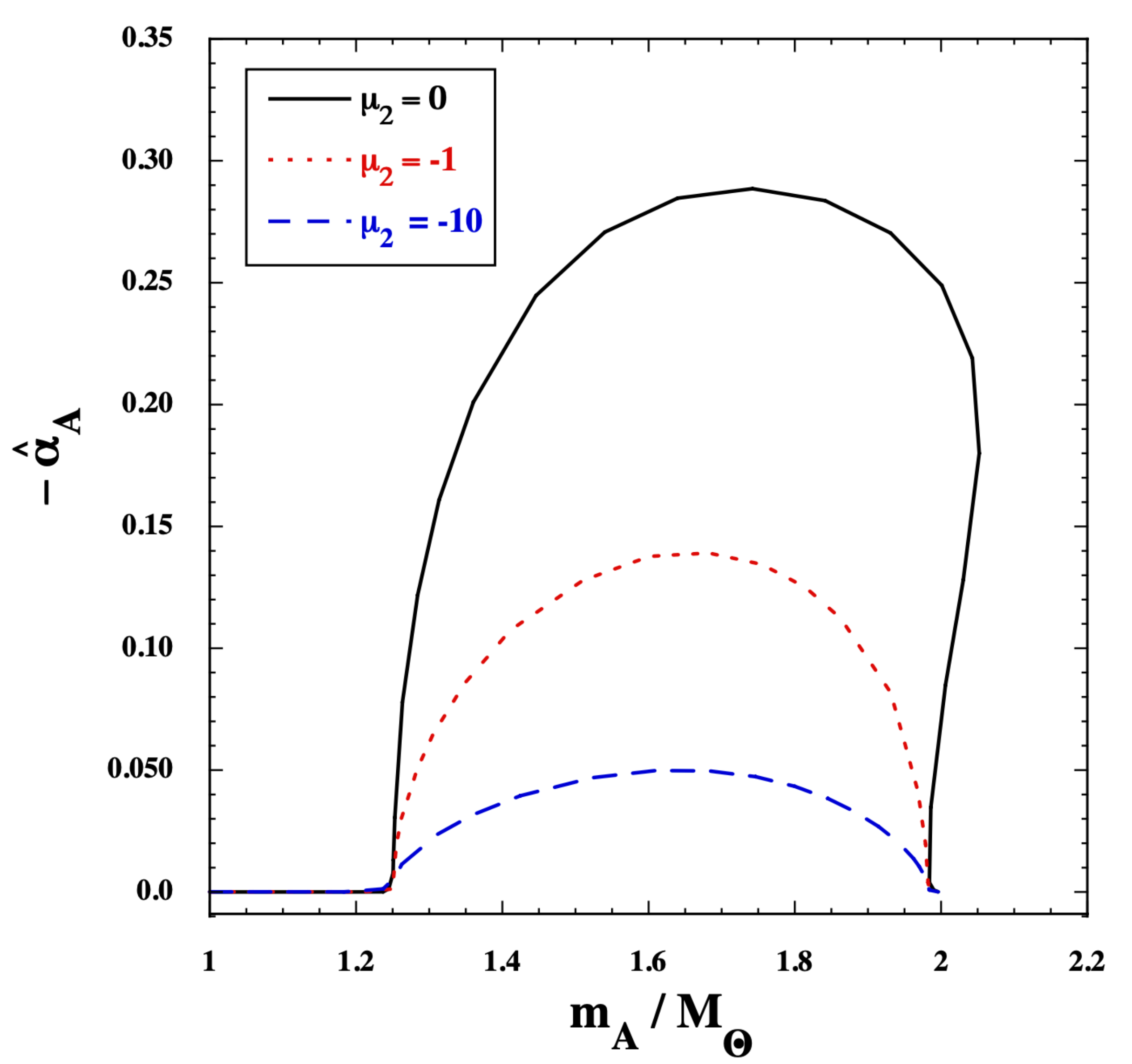}
\end{center}\vspace{-0.5cm}
\caption{\label{fig1}
(Left) Field derivative $|\phi'(r)|$ normalized by $\Mpl/r_0$ in models of 
spontaneous scalarization with $\beta=-5$ for the SLy EOS
with $\rho_c=8 \rho_0$. Each line corresponds to 
$\mu_2=0, -1, -10$, respectively, where $\mu_2$ is normalized 
by $r_0^2/\Mpl^2$.
(Right) $-\hat{\alpha}_A$ versus $m_A/M_{\odot}$ 
for $\beta=-5$ and $\mu_2=0, -1, -10$. 
In this plot, the central matter density
is in the range $\rho_c \leq 12 \rho_0$.
The negative coupling $\mu_2$ leads to the suppression of 
$-\hat{\alpha}_A$ in comparison to standard 
spontaneous scalarization ($\mu_2=0$).
}
\end{figure}

In the left panel of Fig.~\ref{fig1}, we plot the field derivative 
$|\phi'(r)|$ (normalized by $\Mpl/r_0$) with the central density 
$\rho_c=8 \rho_0$ for $\beta=-5$ and $\mu_2=0$ (black line). 
In this case, the radius of NS is $r_s \simeq 0.13 r_0=11.7$\,km 
with the ADM mass $m_A \simeq 1.74 M_{\odot}$, where 
$M_{\odot}$ is the solar mass. 
Deep inside the star ($r \ll r_s$), the scalar field varies according 
to Eq.~(\ref{phicen}), i.e., $\phi'\simeq \beta \phi_c \rho_c (1-3w_c)r/[6 \Mpl^2 F(\phi_c)]<0$ with $w_c=0.239$ and $\phi_c>0$.
For $r \gg r_s$, the solution to $\phi$ is given by Eq.~(\ref{phila2}), i.e., 
$\phi'(r) \simeq m_A \hat{\alpha}_A/[4 \pi F(\phi_0)\Mpl r^2]<0$ 
with $\hat{\alpha}_A<0$. 
As we observe in Fig.~\ref{fig1}, the two solutions smoothly join 
each other around $r=r_s$. 
The scalar field continuously decreases from the central value 
$\phi_c \simeq 0.2835\Mpl$ to the asymptotic value $\phi_0$ 
close to 0 [which is in the range (\ref{phi0})]. 
In this case, the numerical value of $\hat{\alpha}_A$ is 
$-0.29$. The approximate analytic formula (\ref{alI}) gives the value 
$\hat{\alpha}_A \simeq -0.20$, so it is sufficient to 
estimate the order of the scalar charge.

The black line in the right panel of Fig.~\ref{fig1} shows
$-\hat{\alpha}_A$ versus $m_A/M_{\odot}$ for $\beta=-5$ and $\mu_2=0$.
With the central density in the range 
$\rho_c \lesssim 5.25 \rho_0$, the scalar-field solution 
is close to the GR one ($\phi(r)=0$) and hence $|\hat{\alpha}_A|$ is much smaller than 1. 
For $\rho_c \gtrsim 5.25 \rho_0$, which corresponds to 
the ADM mass $m_A \gtrsim 1.25 M_{\odot}$, 
the scalarized branch starts to appear.
With the increase of $\rho_c$, $-\hat{\alpha}_A$ grows to 
reach the maximum value $0.29$ around $\rho_c=8 \rho_0$.
As $\rho_c$ increases further, $-\hat{\alpha}_A$ starts to decrease 
and approaches 0 for $\rho_c \gtrsim 12 \rho_0$. 
This is attributed to the fact that $w_c$ approaches 
$1/3$ in the analytic estimation of Eq.~(\ref{alI}).
For $\mu_2=0$, the observed orbital decay of binary pulsars 
put a stringent limit 
$\beta \geq -4.5$ \cite{Freire:2012mg,Shao:2017gwu,Anderson:2019eay}. 
This bound arises from the scalar radiation of GWs
induced by a large scalar charge. 
Note that the coupling constant $\beta=-5$, which corresponds 
to the black line shown in the right panel of Fig.~\ref{fig1}, 
has been already excluded by binary pulsar measurements.

In the presence of the higher-order derivative term 
$\mu_2 X^2$ with $\mu_2<0$, 
it is possible to reduce $|\hat{\alpha}_A|$. 
The scalar-field equation in the Einstein frame is given by Eq.~(\ref{fieledeqEin}), 
where the right hand-side is $\beta \phi (\hat{\rho}-3 \hat{P})/(2\Mpl^2)$. 
When $\mu_2<0$, the stability condition (\ref{nogho2}) is always satisfied. 
Since the term $(1-3\mu_2 \hat{h} \hat{\phi}'^2)\hat{h}$ gets larger 
for decreasing negative values of $\mu_2$, this leads to the suppression 
of $|\hat{\phi}'|$ especially around the surface of star. 
Then, the scalar field decreases slowly around $r=r_s$. 
Hence we need to choose smaller values of $\phi_c$ to realize  
$\phi_0$ consistent with (\ref{phi0}). 
In the left panel of Fig.~\ref{fig1}, we plot $|\phi'(r)|$ versus $r/r_0$ for 
$\beta=-5$ and $\mu_2=-1, -10$, where $\mu_2$ is normalized by $r_0^2/\Mpl^2$. 
Since $\phi_c$ tends to be smaller for decreasing $\mu_2$, 
the term $\beta \phi_c (\hat{\rho}-3 \hat{P})/(2\Mpl^2)$ is subject to suppression, which results in overall decrease of $|\phi'(r)|$ both inside and outside the NS. 

The suppression of $\phi_c$ induced by the negative coupling constant 
$\mu_2$ leads to the decrease of $|\hat{\alpha}_A|$ through Eq.~(\ref{alI}). 
In the right panel of Fig.~\ref{fig1}, we can confirm that, 
as $\mu_2$ decreases, $|\hat{\alpha}_A|$ gets smaller. 
For $\mu_2=-1$ and $\mu_2=-10$, the maximum numerical values of 
$|\hat{\alpha}_A|$ are 0.14 and 0.05, respectively.
Thus, even when $\beta=-5$, it is possible to realize small values of 
$|\hat{\alpha}_A|$ that can be consistent with binary pulsar constraints. 
For the NS-BH binary with $m_A=1.7 M_{\odot}$ and 
$m_B=2.5 M_{\odot}$, the scalar charge with $|\hat{\alpha}_A|=0.05$ 
corresponds to the ppE parameter $\beta_1$ of order 
$|\beta_1|={\cal O}(10^{-6})$.
If the future GW observations were to put limits on $\beta_1$ 
at this level, it is possible to probe the signature of spontaneous 
scalarization in the coupling ranges $\beta < -4.5$ and $\mu_2<0$.

\subsubsection{Massive theories with $m_s \neq 0$}

Finally, we comment on theories with a nonvanishing scalar-field 
mass $m_s$. If $m_s$ is larger than the typical orbital frequency $\omega \simeq 10^{-13}$~eV during the inspiral phase of compact binaries, 
we showed in Sec.~\ref{GWwavesec} that the gravitational waveforms 
reduce to those in GR. 
For example, let us consider scalarized NSs realized by the coupling functions 
(\ref{symbre}) with the potential (\ref{Vphi}) \cite{Minamitsuji:2022qku}.
In this setup, the scalar field is in a state of symmetry restoration deep inside 
the NS due to the dominance of a positive nonminimal coupling ($\beta>0$) over a negative mass squared of the potential. 
Away from the star, the field settles down at its vacuum expectation value 
$\phi_v=\pi f_B$ with a positive mass squared $m_s^2$. 
In this scenario, the Compton radius $m_s^{-1}$ of the scalar field 
is of order the typical size of NS, i.e., $m_s^{-1}={\cal O}(10\,{\rm km})$, 
so that $m_s={\cal O}(10^{-11}~{\rm eV})$. 
Since $m_s$ is larger than the typical orbital frequency 
$\omega \simeq 10^{-13}$~eV, this model evades constraints from 
the observed gravitational waveform emitted during the inspiral phase.

There are also chameleon theories \cite{Khoury:2003aq,Khoury:2003rn}
in which the effective mass of $\phi$ is large inside the star, 
while the field is light outside the star.  
If the mass $m_s$ outside the NS is smaller than the order
$\omega \simeq 10^{-13}$~eV, there are next-to-leading order 
ppE parameters (\ref{ppE4}) arising from the correction term $m_s^2/\omega^2$ 
besides the leading-order ppE parameters (\ref{ppE2}).
The gravitational waveforms in massive 
BD theories \cite{Alsing:2011er,Berti:2012bp,Sagunski:2017nzb,Liu:2020moh} 
and screened modified gravity 
in the Einstein frame \cite{Zhang:2017srh,Liu:2018sia,Niu:2019ywx}
have been already studied in the literature. 
Our formulation in this paper can accommodate more general 
k-essence theories.

\section{Conclusions}
\label{conclusion}

In this paper, we studied the gravitational waveform emitted during the 
inspiral phase of quasicircular compact binaries in a subclass of 
Horndeski theories. In this class of theories the speed of gravity $c_t$ is equivalent to that of light on the cosmological background, so it automatically evades observational bounds on $c_t$. Our general analysis accommodates not only (massive) BD theories and spontaneous scalarization but also k-essence and cubic derivative interactions. We exploited the PN expansion of a source energy-momentum tensor to derive solutions to the scalar-field perturbation 
from the source to the observer. In the presence of a cubic Galileon coupling 
$\mu_3 X \square \phi$, our formulation is valid for the Vainshtein radius 
$r_V$ at most of order the star radius $r_s$.

In our theory there are no hairy BH solutions known in the literature, but 
nonminimal couplings $G_4(\phi)$ can give rise to NSs endowed 
with scalar charges. We have taken the description of point-like particles 
of the source whose mass $m_I$ depends on the scalar field, 
with $\alpha_I$ defined in Eq.~(\ref{alphaIJ}).  
Due to the presence of nonminimal couplings in the Jordan frame, 
the combination $\hat{\alpha}_I=\alpha_I-g_4/2$ is a quantity 
directly related to the scalar charge, where $g_4$ is defined in Eq.~(\ref{g4J}).
We clarified this point in Sec.~\ref{contheory} by transforming the 
action (\ref{action}) to that in the Einstein frame. 
The nonvanishing values of $\hat{\alpha}_I$ for NSs are crucial 
to probe the signature of modifications of gravity through the 
GW observations.

In Sec.~\ref{weaksec}, we performed the expansion of metric and scalar 
field about a Minkowski background and derived the perturbation 
equations up to second order. We then obtained the quadrupole formula of tensor waves as Eq.~(\ref{thetaij3}), which reduces to the form 
(\ref{thetaijdef}) for a quasicircular orbit of the binary system.
In Sec.~\ref{scapersec}, we showed that the solution to scalar-field 
perturbations up to quadrupole order is given by the sum of a massless 
mode (\ref{phiBso}) and a massive mode (\ref{phimso}). 
The existence of scalar perturbations nonminimally coupled to gravity  
gives rise to breathing ($h_b$) and longitudinal ($h_L$) 
polarizations for the GW field, which are of the forms 
(\ref{hbf}) and (\ref{hLf}) respectively.
We also derived solutions to the TT polarizations of GWs
in the forms (\ref{hp}) and (\ref{ht}).

In Sec.~\ref{waveformsec}, we first discussed the energy loss induced 
by gravitational radiation and computed the time variation of an 
orbital frequency $\omega$. We then derived the gravitational waveform 
in Fourier space under a stationary phase approximation. 
If the scalar-field mass $m_s$ is much smaller than $\omega$, we 
obtained the waveforms of two TT polarizations as
Eqs.~(\ref{hpf}) and (\ref{htf}) under the conditions 
(\ref{delkappa}) and  (\ref{epscon}). 
For $m_s \gg \omega$, the TT modes are practically equivalent to 
those in GR due to the absence of scalar radiation and the exponential 
suppression of fifth forces outside the star. 
In the massless limit $m_s/\omega \to 0$, 
we showed that the leading-order gravitational waveforms reduce to those in 
the ppE formalism, with the ppE parameters (\ref{ppE2}).
If $m_s$ is not very much smaller than $\omega$, there are corrections 
to the ppE parameters given by Eq.~(\ref{ppE4}). 

In Sec.~\ref{contheory}, we computed $\hat{\alpha}_A$ for NSs 
in several concrete theories to confront them with the future observations 
of GWs. In particular, we took into account a higher-order kinetic term 
$\mu_2 X^2$ in $G_2$ for massless BD theories and theories of 
spontaneous scalarization. In BD theories, it is difficult to increase
the scalar charge by the coupling $\mu_2$ due to the appearance of 
ghost or strong coupling problems. 
On the other hand, in theories of spontaneous scalarization, 
the negative coupling $\mu_2$ leads to the suppression of $|\hat{\alpha}_A|$ without inducing ghost instabilities (see the right panel of Fig.~\ref{fig1}).
For $\mu_2=0$, the binary pulsar measurements already put a tight bound 
$\beta \geq -4.5$ on the nonminimal coupling. 
The presence of $\mu_2 X^2$ should make the theory compatible with binary pulsar observations even for $\beta<-4.5$.
It remains to be seen how future events of the NS-BH binary system 
place bounds on the values of $\beta$ and $\mu_2$. 
Finally, we also showed that the recently proposed scalarized NSs 
realized in massive theories with $m_s={\cal O}(10^{-11}~{\rm eV})$ 
and $\beta>0$ give rise to gravitational waveforms similar to those 
in GR. 

It will be of interest to apply our formula of gravitational 
waveforms to the cubic Galileon coupling with 
the Vainshtein radius $r_V \lesssim r_s$.
Moreover, the extension of our analysis to full Horndeski theories
allows us to accommodate more general modified gravity theories 
including the scalar-Gauss-Bonnet coupling \cite{Maselli:2016gxk,Pani:2011xm,Kleihaus:2014lba,Doneva:2017duq,Carson:2020ter,Minamitsuji:2022tze}. 
We leave these topics for future works.

\section*{Acknowledgements}

We are grateful to Luca Amendola, Rampei Kimura, Kei-ichi Maeda, 
Masato Minamitsuji, Atsushi Nishizawa, Hirotada Okawa, 
Hiroki Takeda, David Trestini, and Nicolas Yunes for useful discussions 
and comments. We also thank Tan Liu for answering questions to the paper \cite{Liu:2020moh}.
ST was supported by the Grant-in-Aid for Scientific Research Fund of 
the JSPS Nos.~19K03854 and 22K03642.

\bibliographystyle{mybibstyle}
\bibliography{bib}

\end{document}